\documentclass[a4paper,10pt]{article}
\usepackage[margin=1.in]{geometry}
\usepackage{subfigure,epsfig,amsfonts}
\usepackage{natbib}
\usepackage{amsmath}
\usepackage{amssymb}
\usepackage{amsthm}
\usepackage[utf8]{inputenc}
\usepackage[english]{babel}
\usepackage[nottoc]{tocbibind}
\usepackage{listings} 
\usepackage{enumitem} 
\usepackage{lineno, blindtext} 
\usepackage{xstring} 
\usepackage{longtable}
\usepackage{color}
\usepackage{multicol}
\usepackage{url}

\usepackage{mathtools}

\newcommand{\fontit}[1]{\textit{#1}}

\newcommand{\fontemph}[1]{\emph{#1}}

\newcommand{\name}{EIN}
\newcommand{\EinExp}[2]{\ensuremath{{\left\langle{#1}\right\rangle}_{#2}}}
\newcommand{\EinOp}[3]{\ensuremath{\lambda{}\,#1\EinExp{#2}{#3}}}

\newcommand{\lift}[1]{\textbf{lift}_d({#1})}
 
\newcommand{\einpart}[1]{\frac{\partial}{\partial x_{#1}}\diamond}

\newcommand{\rewriteRule}{\xrightarrow[rule]{}}
\newcommand{\rewriteNorm}{\rewriteRule^*}

\newcommand{\TensorTy}[1]{\mathbf{Ten}[#1]}
\newcommand{\FieldTy}[2]{\mathbf{Fld}(#1)[#2]}
\newcommand{\ImageTy}[2]{\mathbf{Img}(#1)[#2]}
\newcommand{\KernelTy}{\mathbf{Krn}}
\newcommand{\genTTyB}{\ensuremath{\mathcal{T}}}      
\newcommand{\genFTyB}{\ensuremath{\mathcal{F}^{d}}}  %
\newcommand{\genTTyA}[1]{\ensuremath{(#1)\genTTyB{}}}      
\newcommand{\genFTyA}[1]{\ensuremath{(#1)\genFTyB{}}}  

\newtheorem{theorem}{Theorem}[section]
\newtheorem{lemma}[theorem]{Lemma}
\newenvironment{definition}[1][Definition]{\begin{trivlist}
\item[\hskip \labelsep {\bfseries #1}]}{\end{trivlist}}

\newcommand{\secref}[1]{Section~\ref{#1}}
\newcommand{\tblref}[1]{Table~\ref{#1}}
\newcommand{\figref}[1]{Figure~\ref{#1}}
\newcommand{\eqqref}[1]{Equation~\ref{#1}}
\newcommand{\thmref}[1]{Theorem~\ref{#1}}
\newcommand{\lemref}[1]{Lemma~\ref{#1}}
\newcommand{\eg}{{\em e.g.}}

\newcommand{\etc}{{\em etc.\/}}

\newcommand{\role}{r\^{o}le}

\definecolor{Black}{rgb}{0.0,0.0,0.0}

\newcommand{\cdColor}{Black}

\newcommand{\LDB}{\ensuremath{[\mskip -3mu [}}
\newcommand{\RDB}{\ensuremath{]\mskip -3mu ]}}
\newcommand{\dom}{\ensuremath{\mathrm{dom}}}

\newcommand{\SET}[1]{\ensuremath{\{#1\}}}

\newcommand{\finmap}{\mathrel{\stackrel{\textrm{fin}}{\rightarrow}}}

\definecolor{ngray}{rgb}{0.5,0.5,0.5}

\lstdefinelanguage{Diderot}{%
  morekeywords={%
    bool,%
    die,%
    else,%
    false,field,foreach,%
identity,if,image,load,in,initially,input,int,%
    fem,
    kernel,%
    nan,new,%
    output,%
    real,%
    function,stabilize,strand,string,%
    inside,tensor,true,%
    update,%
    vec2,vec3,vec4,%
    zeros},
  sensitive,%
  morecomment=[s]{/*}{*/},%
  morecomment=[l]//,
  morestring=[b]"}%

\newcommand{\RuleGroupA}[2]{
 \IfEqCase{#1}{
     {1}{\IfEqCase{#2}{
          {0}{R35}{1}{A1}
          {2}{ \mathcal{E}_{ijk} \mathcal{E}_{ilm} 
                  \rewriteRule{}\delta_{jl}\delta_{km} -\delta_{jm}\delta_{kl} }
          {3}{\mathcal{E}_{ijk} \mathcal{E}_{ilm}}
         {4}{\delta_{jl}\delta_{km} -\delta_{jm}\delta_{kl}}
          }}
    {3}{\IfEqCase{#2}{ {0}{R34}{1}{A3}
          {2}{  \mathcal{E}_{ijk} (V_\alpha \circledast h^{jk}) \rewriteRule{} \lift{0}  }
          {3}{\mathcal{E}_{ijk} (V_\alpha \circledast h^{jk})}{4}{\lift{0}} }}
    {4}{\IfEqCase{#2}{{0}{R33}{1}{A4}
          {2}{  \mathcal{E}_{ijk}  \einpart{i,j}  e_1 \rewriteRule{} \lift{0}   }
          {3}{\mathcal{E}_{ijk}  \einpart{i,j}  e_1}{4}{\lift{0}}  }}
    {5}{\IfEqCase{#2}{{0}{R36}{1}{A5} {2}{\delta_{ij} T_{j} \rewriteRule{} T_i  }
         {3}{\delta_{ij} T_{j} }{4}{ T_i } }}
    {6}{\IfEqCase{#2}{{0}{R37}{1}{A6}{2}{ \delta_{ij} F_{j} \rewriteRule{} F_{i} } {3}{ \delta_{ij} F_{j} }{4}{ F_i  }}}
    {7}{\IfEqCase{#2}{{0}{R40}{1}{A7}
         {2}{ \delta_{ij} \einpart{j} e_1 \rewriteRule{}  \einpart{i} (e_1)  }
         {3}{ \delta_{ij} \einpart{j} (e_1) }{4}{\einpart{i} (e_1)  } }}
   {8}{\IfEqCase{#2}{{0}{R38}{1}{A8}
         {2}{ \delta_{ij} V \circledast H ^{\delta_{cj}} \rewriteRule{}    V \circledast H^{\delta_{ci}} }
         {3}{ \delta_{ij} V \circledast H ^{\delta_{cj}} }{4}{   V \circledast H^{\delta_{ci}} }}}  
            {9}{\IfEqCase{#2}{{0}{R39}{1}{A8}
         {2}{ \delta_{ij} V \circledast H ^{\delta_{cj}}(x) \rewriteRule{}    V \circledast H^{\delta_{ci}}(x) }
         {3}{ \delta_{ij} V \circledast H ^{\delta_{cj}} (x)}{4}{   V \circledast H^{\delta_{ci}} (x)}}}  
}}

\newcommand{\RuleGroupB}[2]{
 \IfEqCase{#1}{
    {1}{\IfEqCase{#2}{{0}{R1}{1}{B1}
        {2}{    ( e_1 \odot_n e_2)@x \rewriteRule{}   (e_1 @x) \odot_n (e_2 @x)  }
        {3} {( e_1 \odot_n e_2)@x}{4}{ (e_1 @x) \odot_n (e_2@x) }}}
    {2}{\IfEqCase{#2}{{0}{R2} {1}{B2}
        {2}{  ( e_0 \odot_2 e_1) @x \rewriteRule{} (e_0 @x )\odot_2(e_1@x)   }
         {3} {( e_0\odot_2 e_1) @x }{4}{(e_0 @x)\odot_2 (e_1@x) }}}         
     {3}{\IfEqCase{#2}{{0}{R3}{1}{B3}{2}{    ( \odot_1 e_1)@x \rewriteRule{} \odot_1  (e_1@x)  }
        {3} { (\odot_1 e_1)@x}{4}{  \odot_1   (e_1@x) }}}
      {4}{\IfEqCase{#2}{{0}{R4}{1}{B4}{2}{ ( \sum\limits_{i=1}^{n}  e_1) @x \rewriteRule{} \sum\limits_{i=1}^{n} (e_1@x) }
             {3}{( \sum\limits_{i=1}^{n}  e_1) @x}{4}{ \sum\limits_{i=1}^{n}  (e_1@x)}}}
     {5}{\IfEqCase{#2}{{0}{R5}{1}{B5}{2}{ (\chi) @x \rewriteRule{}  \chi}{3} {(\chi) @x}{4}{\chi}}}                                 
      }}

\newcommand{\RuleGroupC}[2]{ \IfEqCase{#1}{
     {1}{
         \IfEqCase{#2}{
             {0}{R20}
             {1}{C1}
             {2}{ \frac{\partial}{\partial x_i} \lift{e_1}\rewriteRule{}  0  }
             {3}{ \frac{\partial}{\partial x_i} \text{Lift}(e_1) }
             {4} {0}  
             }
         }
    {2}{ 
        \IfEqCase{#2}{
            {0}{R20}
            {1}{C2}
            {2}{ \frac{\partial}{\partial x_i}  \chi \rewriteRule{} 0}      
            {3}{  \frac{\partial}{\partial x_i}    }
            {4}{ 0} 
            }
        }
    {3}{\IfEqCase{#2}{{0}{R19}{1}{C3}
        {2}{ \frac{\partial}{\partial x_i}  \sum\limits_{v=1}^{n} e_1 \rewriteRule{} \sum\limits_{v=1}^{n} (\frac{\partial}{\partial x_i}  e_1) }
         {3}{ \frac{\partial}{\partial x_i}  \sum\limits_{v=1}^{n}  e_1  }{4}{ \sum\limits_{v=1}^{n}(\frac{\partial}{\partial x_i}  e_1)} }}
    {5}{\IfEqCase{#2}{ {0}{R21} {1}{C5}
        {2}{ {\einpart{i} (V_{\alpha} \circledast H^{\nu} )} \rewriteRule{} (V_\alpha \circledast h^{i \nu}) }
        {3}{\einpart{i} (V_{\alpha} \circledast H^{\nu} )}{4}{ (V_\alpha \circledast H^{i \nu})}}}
    {6}{ \IfEqCase{#2}{{0}{R8}{1}{C6}
       {2}{ \einpart{i}  (\sqrt {e_1}) \rewriteRule{} \lift{1/2} *\frac{ \einpart{i}  e_1}{\sqrt {e_1}} }
       {3}{ \einpart{i}  (\sqrt {e_1}) }{4}{ \lift{1/2} *\frac{ \einpart{i}  e}{\sqrt {e_1}} } } }
    {7}{\IfEqCase{#2}{{0}{R9}{1}{C7}
        {2}{  \einpart{i}  (\textbf{cosine}(e_1)) \rewriteRule{} (-\textbf{sine}(e_1))*( \einpart{i}  e_1) }
        {3}{ \einpart{i}  (\textbf{cosine}(e_1)) }{4}{(-\textbf{sine}(e_1))*( \einpart{i}  e_1)} } } 
     {8}{\IfEqCase{#2}{{0}{R10}{1}{C8}
        {2}{ \einpart{i}  (\textbf{sine} (e_1)) \rewriteRule{} (\textbf{cosine} (e_1))*( \einpart{i}  e_1) }
        {3}{\einpart{i}  (\textbf{sine} (e_1)) }{4}{(\textbf{cosine} (e_1))*( \einpart{i}  e_1) }}}
    {9}{\IfEqCase{#2}{{0}{R12}{1}{C9}
        {2}{  \einpart{i}  (\textbf{arccosine}(e_1))\rewriteRule{} (\frac{-\lift{1}}{\sqrt{(\lift{1}-(e*e)}} )*(\einpart{i} e_1)}
        {3}{\einpart{i}  (\textbf{arccosine} (e_1)) }{4}{(\frac{-\lift{1}}{\sqrt{(\lift{1}-(e*e)}} )*( \einpart{i}  e_1) }} }
    {10}{ \IfEqCase{#2}{{0}{R13} {1}{C10}
       {2}{ \einpart{i}  (\textbf{arcsine}(e_1)) \rewriteRule{} (\frac{\lift{1}}{\sqrt{(\lift{1}-(e*e)}} )* (\einpart{i}  e_1) }
       {3}{ \einpart{i}  (\textbf{arcsine}(e_1))}{4}{ (\frac{\lift{1}}{\sqrt{(\lift{1}-(e*e)}} )* (\einpart{i}  e_1) }} }  
   {11}{
         \IfEqCase{#2}{ {0}{R7}{1}{C11}
       {2}{ \einpart{i}  (\frac{e_1}{e_2}) \rewriteRule{}  \frac{ (\einpart{i}  e_1) e_2 -e_1 (\einpart{i}  e_2)}{e_2^2} }
       {3}{ \einpart{i}  (\frac{e_1}{e_2})}
       {4}{  \frac{ (\einpart{i}  e_1) e_2 -e_1 (\einpart{i}  e_2)}{e_2^2} }}}  
    {14}{ \IfEqCase{#2}{{0}{R6}{1}{C14}
       {2}{ \einpart{i}  (e_1 * e_2) \rewriteRule{} e_1 (\einpart{i}  e_2) +e_2 (\einpart{i}  e_1) }
       {3}{ \einpart{i}  (e_1 * e_2) }{4}{e_1 \einpart{i}  e_2 +e_2 \einpart{i}  e_1 }}}  
     {15}{ \IfEqCase{#2}{{0}{R18}{1}{C15}
       {2}{ \einpart{i} (- e_1) \rewriteRule{} -  (\einpart{i}  e_1) }
       {3}{ \einpart{i} (- e_1)  }{4}{-  (\einpart{i}  e_1 )}}} 
   {16}{ \IfEqCase{#2}{{0}{R17}{1}{C16}
       {2}{ \einpart{i} (e_1 \odot e_2) \rewriteRule{}   (\einpart{i}  e_1)  \odot( \einpart{i} e_2) }
       {3}{ \einpart{i} (e_1 \odot e_2) }{4}{(\einpart{i}  e_1)  \odot( \einpart{i} e_2)}}}
  {18}{ \IfEqCase{#2}{{0}{R11}{1}{C18}
       {2}{  \einpart{i}  (\textbf{tangent} (e_1)) 
              \rewriteRule{}   \frac{ \einpart{i}  e}{\textbf{cosine}(e_1)* \textbf{cosine}(e_1)} }
       {3}{ \einpart{i}  (\textbf{tangent} (e_1)) }{4}{\frac{ \einpart{i}  e}{\textbf{cosine}(e_1)* \textbf{cosine}(e_1)} }}}
  {19}{ \IfEqCase{#2}{{0}{R14}{1}{C19}
        {2}{ \einpart{i} (\textbf{arctangent}(e_1))
             \rewriteRule{}\frac{\lift{1}}{\lift{1}+(e_1*e_1)}* (\einpart{i}  e_1)  }
       {3}{\einpart{i} (\textbf{arctangent}(e_1)) }
       {4}{\frac{1}{1+(e*e)}* (\einpart{i}  e_1) }}}
 {20}{\IfEqCase{#2}{{0}{R15}{1}{C20} 
       {2}{\einpart{i} (\textbf{exp}(e_1)) \rewriteRule{} \textbf{exp}(e_1)* (\einpart{i}  e_1)  } 
       {3}{ \einpart{i} (\textbf{exp}(e_1)) }{4}{\textbf{exp}(e_1)* (\einpart{i}  e_1) }  } }  
  {21}{\IfEqCase{#2}{{0}{R16}{1}{C21} 
       {2}{\einpart{i}  (e_1^n) \rewriteRule{} \lift{n}* e_1^{n-1} * (\einpart{i}  e_1)   } 
       {3}{\einpart{i}  (e_1^n) }{4}{\lift{n}* e_1^{n-1} * (\einpart{i}  e_1)  }  } }   
        {22}{
            \IfEqCase{#2}{{0}{R42}{1}{C22} 
               {2}{\einpart{\alpha}  \einpart{\beta}  e_1 \rewriteRule{} \einpart{\beta \alpha}  e_1 } 
               {3}{\einpart{\alpha}  \einpart{\beta}  e_1 }{4}{\einpart{\beta \alpha}  e_1 }  } }                 
        }
    }

\newcommand{\RuleGroupD}[2]{
 \IfEqCase{#1}{
    {1}{\IfEqCase{#2}{{0}{R23} {1}{D1}{2}{-0 \rewriteRule{} 0}{3}{-0}{4}{0}}}
    {2}{\IfEqCase{#2}{{0}{R30}{1}{D2}{2}{0+e_1,e_1+0 \rewriteRule{} e_1 }{3}{0+e_1,e_1+0 }{4}{e_1}}}
    {3}{\IfEqCase{#2}{{0}{R24}{1}{D3}{2}{ e_1-0 \rewriteRule{} e_1}{3}{ e_1-0}{4}{e_1}}}
    {4}{\IfEqCase{#2}{{0}{R25}{1}{D4}{2}{ 0-e_1 \rewriteRule{} -e_1 }{3}{ 0-e_1 }{4}{ -e_1 }}}
    {5}{\IfEqCase{#2}{{0}{R26}{1}{D5}{2}{\frac{0}{e_1} \rewriteRule{} 0}{3}{\frac{0}{e_1}}{4}{ 0}}}
    {6}{\IfEqCase{#2}{{0}{R31}{1}{D6}{2}{0e,e0 \rewriteRule{} 0}{3}{0e,e0}{4}{0}}}
    {8}{\IfEqCase{#2}{{0}{R22}{1}{D7}{2}{--e_1 \rewriteRule{} e_1}{3}{--e_1}{4}{e_1}}}      
}}

\newcommand{\RuleGroupE}[2]{
 \IfEqCase{#1}{
    {1}{\IfEqCase{#2}{{0}{R27}{1}{E1}{2}{\frac{\frac{ e_1 }{ e_2} }{ e_3}  \rewriteRule{}  \frac{e_1} { e_2 e_3}  }
        {3}{\frac{\frac{ e_1 }{ e_2} }{ e_3}  }{4}{ \frac{e_1} { e_2 e_3}}}}
    {2}{\IfEqCase{#2}{{0}{R28}{1}{E2}{2}{\frac{e_1 } {\frac{ e_2 }{ e_3} }      \rewriteRule{}  \frac{e_1 e_3} { e_2}}
        {3}{\frac{e_1 } {\frac{ e_2 }{ e_3} } }{4}{ \frac{e_1 e_3} { e_2} }}}
    {4}{\IfEqCase{#2}{{0}{R29}{1}{E4}
        {2}{ \frac{\frac{e_1}{e_2 }}   {\frac{e_3 }{ e_4} } 
             \rewriteRule{}  \frac{e_1 e_4} {e_2 e_3}}
        {3}{ \frac{\frac{e_1}{e_2 }}   {\frac{e_3 }{ e_4} } }
        {4}{  \frac{e_1 e_4} {e_2 e_3}}}}
   {5}{ \IfEqCase{#2}{{0}{R41} {1}{E5}{2}{\sum (s e_1) \rewriteRule{} s\sum e_1  }
       {3}{\sum (s e_1)}{4}{s \sum e_1  }}}
   {6}{\IfEqCase{#2}{{0}{R32}{1}{E6}{2}{ \sqrt{(e_1)} * \sqrt{(e_1)}  \rewriteRule{}  e_1 }
       {3}{ \sqrt{(e_1)} * \sqrt{(e_1)} }{4}{ e_1 }}}
    }}

\newcommand{\RuleGroupG}[2]{
 \IfEqCase{#1}{
      {7}{\IfEqCase{#2}{{0}{R41}{1}{C8}
            {2}{$e \rewriteRule{} e' $\\ $\spcA \line(1,0){100} $\\$\spcC \sum e_1 \rewriteRule{} \sum e'$ }}}
      {1}{\IfEqCase{#2}{{0}{R43}{1}{C8}
            {2}{ $e \rewriteRule{} e' $\\$\spcA \line(1,0){100} $\\  $\spcC \zeta e_1 \rewriteRule{}  \zeta e'$}}}
      {2}{\IfEqCase{#2}{{0}{R44}{1}{C8}
            {2}{ $e \rewriteRule{} e' $\\ $\spcA \line(1,0){100} $\\  $\spcC \kappa e_1 \rewriteRule{}  \kappa e'$  }}}
      {3}{\IfEqCase{#2}{{0}{R45}{1}{C8}
            {2}{ $ e_1 \rewriteRule{} e'  \phantom{...} {\&} \phantom{...}x \rewriteRule{} x' $ \\
             $\spcA  \line(1,0){80}$  \\ $\spcA  e_1 @x  \rewriteRule{}  e'@x'  $}}}
    {4}{\IfEqCase{#2}{{0}{R46}{1}{C8}
        {2}{$ e_1 \rewriteRule{} e_1'  \phantom{...} {\&} \phantom{...} e_2 \rewriteRule{} e_2' $ \\
            $ \spcA \line(1,0){100}  $\\ $\spcC {e_1  e_2} \rewriteRule{} {e_1'  e_2'} $}}}
    {5}{\IfEqCase{#2}{{0}{R47}{1}{C8}
        {2}{$e_1 \rewriteRule{} e_1'  \phantom{...} {\&} \phantom{...} e_2\rewriteRule{} e_2' $\\
            $\spcA \line(1,0){100}$ \\ $\spcC  e_1 \Omega e_2  \rewriteRule{}  e_1' \Omega e_2'$}}}
    {6}{\IfEqCase{#2}{{0}{R48}{1}{C8}
        {2}{ $ e_1 \rewriteRule{} e_1'  \phantom{...} {\&} \phantom{...} es \rewriteRule{} es'$\\
        $\spcA  \line(1,0){100}$\\ $\spcC  e_1*es \rewriteRule{} e_1' *es'$} }}
     }}

\newcommand{\RuleGroupGInd}[1]{
 \IfEqCase{#1}{
      {7}{R41&G7&$e \rewriteRule{} e' $\\
           && $\line(1,0){100} $\\
            && $\sum e_1 \rewriteRule{} \sum e'$ }
      {1} {R42&G1&           $e \rewriteRule{} e' $\\
           && $\line(1,0){100} $\\             &&$\zeta e_1 \rewriteRule{}  \zeta e'$}
      {2}{R43&G2&           $e \rewriteRule{} e' $\\
           && $\line(1,0){100} $\\            &&$\kappa e_1 \rewriteRule{}  \kappa e'$  }
        {3}{R44 &G3& $ e_1 \rewriteRule{} e'  \phantom{...} {\&} \phantom{...}x \rewriteRule{} x' $ \\
             && $ \line(1,0){80}$  \\
             && $  e_1 @x  \rewriteRule{}  e'@x'  $}
        {4}{R45&G4&$ e_1 \rewriteRule{} e_1'  \phantom{...} {\&} \phantom{...} e_2 \rewriteRule{} e_2' $ \\
 &&$ \line(1,0){100}  $\\
   &&${e_1  e_2} \rewriteRule{} {e_1'  e_2'} $}
           {5}{46&G5&$e_1 \rewriteRule{} e_1'  \phantom{...} {\&} \phantom{...} e_2\rewriteRule{} e_2' $\\
&&$\line(1,0){100}$ \\
&&$ e_1 \Omega e_2  \rewriteRule{}  e_1' \Omega e_2'$}
      {6}{R47&G6&  $ e_1 \rewriteRule{} e_1'  \phantom{...} {\&} \phantom{...} es \rewriteRule{} es'$\\
&&$ \line(1,0){100}$\\
 && $ e_1*es \rewriteRule{} e_1' *es'$} 
}}

\newcommand{\ruleNum}[2]{\hspace{-0.3em}\IfEqCase{#1}{
    {A}{\RuleGroupA{#2}{0}}{B}{\RuleGroupB{#2}{0}}{C}{\RuleGroupC{#2}{0}}
    {D}{\RuleGroupD{#2}{0}}{E}{\RuleGroupE{#2}{0}}{G}{\RuleGroupG{#2}{0}}}
}
\newcommand{\ruleCmd}[2]{\IfEqCase{#1}{
    {A}{A{#2}}{B}{B{#2}}{C}{C{#2}}
    {D}{D{#2}}{E}{E{#2}}{G}{G{#2}}}
}
\newcommand{\ruleEin}[2]{
\IfEqCase{#1}{
    {A}{\RuleGroupA{#2}{2}}
    {B}{\RuleGroupB{#2}{2}}
    {C}{\RuleGroupC{#2}{2}}
    {D}{\RuleGroupD{#2}{2}}
    {E}{\RuleGroupE{#2}{2}}
    {G}{\RuleGroupG{#2}{2}}
    }
}
\newcommand{\ruleSrc}[2]{ \IfEqCase{#1}{
    {A}{\RuleGroupA{#2}{3}}
    {B}{\RuleGroupB{#2}{3}}
    {C}{\RuleGroupC{#2}{3}}
    {D}{\RuleGroupD{#2}{3}}{E}{\RuleGroupE{#2}{3}}{G}{\RuleGroupG{#2}{3}}}
}
\newcommand{\ruleTar}[2]{ \IfEqCase{#1}{
    {A}{\RuleGroupA{#2}{4}}{B}{\RuleGroupB{#2}{4}}{C}{\RuleGroupC{#2}{4}}
    {D}{\RuleGroupD{#2}{4}}{E}{\RuleGroupE{#2}{4}}{G}{\RuleGroupG{#2}{4}}}
}

\newcommand{\spc}[0]{ \phantom{*}}
\newcommand{\spcA}[0]{\spc\spc}
\newcommand{\spcB}[0]{\spcA\spc}
\newcommand{\spcC}[0]{\spcB\spc}
\newcommand{\spcD}[0]{\spcC\spc}
\newcommand{\spcE}[0]{\spcD\spc}	 
	
\newcommand{\basen}{b_1 \dots b_n}

\newcommand{\thmNameSize}{P}
\newcommand{\thmNameN}{Q}
\newcommand{\thmNameL}{M}
\newcommand{\thmNameV}{V}
\newcommand{\thmNameT}{T}
\newcommand{\OkJud}[2]{{#1} \vdash {#2} \; \textbf{ok}}
\newcommand{\TyHead}[2]{\newline \textbf{Case }
\ruleNum{#1}{#2}.$\ruleEin{#1}{#2}$}
\newcommand{\donecaseG}[3]{\spc {\textcolor{\cdColor}{\texttt{{#3}(\ruleNum{#1}{#2})  OK}}}}
\newcommand{\donecaseT}[2]{\donecaseG{#1}{#2}{\thmNameT}}
\newcommand{\donecaseV}[2]{\donecaseG{#1}{#2}{\thmNameV}}
\newcommand{\donecaseP}[2]{\donecaseG{#1}{#2}{\thmNameSize}}
\newcommand{\donecaseM}[2]{\donecaseG{#1}{#2}{\thmNameL}}
\newcommand{\sameRuleTyG}[4]{
\TyHead{#1}{#2}\\
\spcC  Similar approach to \ruleNum{#3}{#4}
}
\newcommand{\sameRuleTyT}[4]{\sameRuleTyG{#1}{#2}{#3}{#4} \donecaseT{#1}{#2} }
\newcommand{\sameRuleTyV}[4]{\sameRuleTyG{#1}{#2}{#3}{#4} \donecaseV{#1}{#2} }
\newcommand{\sameRuleTyP}[4]{\sameRuleTyG{#1}{#2}{#3}{#4} \donecaseP{#1}{#2} } 
 
\newcommand{\TyGiven}[2]{Given that ${#1}$}
\newcommand{\TyApply}[2]{${#2}$ by ${#1}$}
\newcommand{\TyBy}[3]{${#2} = {#3}$ by ${#1}$}
\newcommand{\VApply}[2]{ $\Psi, \rho \vdash{#2}$ by ${#1}$}

\newcommand{\einTyB}[1]{({#1})\tau_0}
\newcommand{\einTyA}{\tau}

\newcounter{typecnt}

\newcommand{\TyJud}[3]{{#1} \vdash {#2} : {#3}}
\newcommand{\EvalJud}[3]{{#1} \vdash {#2} \Downarrow {#3}}

\newcommand{\TyEnv}{\Gamma,\sigma}
\newcommand{\linegam}[2]{\TyJud{\TyEnv}{#1}{#2}}
\newcommand{\linegamT}[1]{\TyJud{\TyEnv}{#1}{\genTTyA{\sigma}}}
\newcommand{\linegamR}[1]{\TyJud{\TyEnv}{#1}{\genTTyA{}}}
\newcommand{\linegamF}[1]{\TyJud{\TyEnv}{#1}{\genFTyA{\sigma}}}
\newcommand{\linegamD}[1]{\TyJud{\TyEnv[i \mapsto (1,d)]}{#1}{\genFTyA{i}}}
\newcommand{\linegamS}[1]{\TyJud{\TyEnv}{#1}{\genFTyA{}}}
\newcommand{\linegamA}[1]{\TyJud{\TyEnv}{#1}{\einTyB{\sigma}}}
\newcommand{\linegamAS}[1]{\TyJud{\TyEnv}{#1}{\einTyB{}}}

\newcommand{\inferruleLayer}[3]{\inferrule{\inferrule{#1}{#2}}{#3}}
\newcommand{\TyFind}[1]{Find $\Gamma, \sigma \vdash ({#1})$\\ }

\newcommand{\TyClaimGen}[1]{Assign generic type $\linegam{#1}{\tau}$}
\newcommand{\TybodyCase}[3]{
 We will do a case analysis on the structure on the 
left-hand-side \\
\spcB where {#1}=\{{#2}$\mid${#3}\}. \\ 
First  we will prove T(d) for {#1}={#2} then {#1}={#3}.\\
}
\newcommand{\TybodyProbe}{
This type of structure inside a probe operation results in a tensor type.\\ 
}
\newcommand{\TybodyDeriv}[1]{
This type of structure inside a derivative operation results in a field type. \\
 Given the subterm: $\Gamma, \sigma /\ i \vdash  {#1}:\genFTyA{\sigma /\ i}$ \\ 
 \spcB then by $\judgeDeriv$ we know it's derivative 
$\Gamma, \sigma \vdash  \nabla_i \diamond ({#1}):\genFTyA{\sigma}$\\
}
\newcommand{\TybodyDerivS}[1]{
 This type of structure inside a derivative operation results in a field type \\
 \spcB and 
 the ${#1}$ term results in a scalar.\\
Claim: $\Gamma_{} \vdash  {#1}:\genFTyA{}$ then 
\TyApply{\judgeDeriv}{\Gamma_{i} \vdash  \nabla_i \diamond ({#1}):\genFTyA{i}}
}
\newcommand{\TybodySrcBB}[4]{ $\linegam{\ruleSrc{#1}{#2}}{#3}$ \spcB ({ $#4$}) }

\newcommand{\TyLastFind}{From that we can make the RHS derivations.\\}
\newcommand{\TyLHS}[3]{
The LHS has the following type. \\
\spcB $\Gamma, \sigma  \vdash\ruleSrc{#1}{#2}:${#3}\\
We want to show that the RHS has the same type.\\
\spcB $\Gamma, \sigma \vdash\ruleTar{#1}{#2}$:{#3}.\\
The type derivation for the LHS is the following structure.\\}
\newcommand{\valcore}[1]{#1}
\newcommand{\valopU}[2]{#1(#2)}
\newcommand{\valopB}[3]{#2 #1 #3}
\newcommand{\valReal}[1]{\valcore{\fontit{Real}(#1)}}
\newcommand{\valTensor}[2]{\valcore{\fontit{Tensor}[{#1} \cdot {#2}]}}

\newcommand{\valP}[1]{\valcore{K_{#1}}}
\newcommand{\valE}[1]{\valcore{E_{#1}}}
\newcommand{\valProbe}[2]{\valcore{\fontit{Probe}({#1})[{#2}] }}
\newcommand{\valNo}[2]{\newline \textbf{Case }
\ruleNum{#1}{#2}.$\ruleEin{#1}{#2}$ \newline \spcB Value representation not supported}

\newcommand{\valTarM}[2]{
\text{We need to show that }\ruleTar{#1}{#2}\text{ evaluates to }v.\\ 
}

\newcommand{\valHeading}[2]{
\spc\text{Claim }\ruleSrc{#1}{#2}\text{ evaluates to }v.\\
\spc \text{We need to define }v.\\
}

\newcommand{\valHeadingR}[2]{
\valHeading{#1}{#2} 
\spc  \text{Assume that } \ruleTar{#1}{#2}  \Downarrow v'
}

\newcommand{\valProbeBinRule}[1]{ 
 $\infer
{ \EvalJud{\Psi, \rho}{e_1 }{ v_1} \spc  \EvalJud{\Psi, \rho}{e_2}{ v_2} \spc \odot_2 =+\mid - \mid * \mid /\ }
{ \EvalJud{\Psi, \rho}{(e_1 \odot_2  e_2) @ x }{
\valopB{\odot_2}{\valProbe{v_1}{x} }{ \valProbe{v_2}{x} } }  } $
}

\newcommand{\valEndProse}[2]{
\spc \text{The last step leads to }\ruleTar{#1}{#2}\Downarrow v\\
\donecaseV{#1}{#2}
}

\newcommand{\TyJudge}{[\text{TYJUD}}
\newcommand{\judgeBase}{\TyJudge_1] }
\newcommand{\judgeConv}{\TyJudge_2] }
\newcommand{\judgeSum}{\TyJudge_3] }
\newcommand{\judgeDeriv}{\TyJudge_4] }
\newcommand{\judgeDel}{\TyJudge_5] }

\newcommand{\judgeProdDel}{\TyJudge_5] }
\newcommand{\judgeProdEps}{\TyJudge_6] }
\newcommand{\judgeProbe}{\TyJudge_7] }
\newcommand{\judgeLift}{\TyJudge_8] }
\newcommand{\judgeM}{\TyJudge_9] }
\newcommand{\judgeNeg}{\TyJudge_{10}] }
\newcommand{\judgeAdd}{\TyJudge_{10}] }
\newcommand{\judgeProd}{\TyJudge_{11}] }
\newcommand{\judgeDiv}{\TyJudge_{12}] }
\newcommand{\TyInv}[1]{[\text{TYINV}_{#1}]}
\newcommand{\invSum}{\TyInv{3}}
\newcommand{\invDeriv}{\TyInv{4}}

\newcommand{\invProbe}{\TyInv{7}}
\newcommand{\invLift}{\TyInv{8}}
\newcommand{\invM}{\TyInv{9}}
\newcommand{\invNeg}{\TyInv{10}}
\newcommand{\invAdd}{\TyInv{10}}
\newcommand{\invProd}{\TyInv{11}}
\newcommand{\invDiv}{\TyInv{12}}

\newcommand{\invgamN}[3]{  &$\linegam{#1}{#2} $&$\mapsto  {#3}$}

\newcommand{\valJudgmentName}[1]{[\text{VALJUD}_{#1}]}
\newcommand{\valbaseC}{\valJudgmentName{1}}
\newcommand{\valbaseT}{\valJudgmentName{2}}
\newcommand{\vallift}{\valJudgmentName{3}}
\newcommand{\valunary}{\valJudgmentName{4}}
\newcommand{\valbinary}{\valJudgmentName{5}}

\newcommand{\valprobe}{\valJudgmentName{6}}
\newcommand{\valdel}{\valJudgmentName{7}}
\newcommand{\valeps}{\valJudgmentName{7}}

\newcounter{rulecnt}

\newcommand{\src}[1]{\text{source}({#1})}
\newcommand{\tar}[1]{\text{target}({#1})}

\newcommand{\size}[0]{\mathcal{S}}

\newcommand{\sizeof}[1]{\LDB{#1}\RDB}
\newcommand{\sizee}[0]{\sizeof{e}}
\newcommand{\sizea}[0]{\sizeof{e_1}}
\newcommand{\sizeb}[0]{\sizeof{e_2}}

\newcommand{\thmTlastline}[1]{ 
\spcC  $\sizeof{\src{d}}> \sizeof{\tar{d}} $\spcB {#1}\\
\spcC\thmNameSize(d)}

\newcommand{\thmTComp}[4]{$\begin{array}{lll}\sizeof{{#1}} &=&  {#2} \\ & >& {#3} \\ & =& \sizeof{#4} \end{array}$\\
}
\newcommand{\thmTCompBy}[5]{\spcB $\begin{array}{lll}\sizeof{{#1}} &=&  {#2} \\ & >& {#3} \\ & =& \sizeof{#4} \end{array}$\\
}
\newcommand{\thmTLong}[5]{
    {\thmTCompBy
        {\ruleSrc{#1}{#2}}
        {{#3}}
        {{#4}}
        {\ruleTar{#1}{#2}}
        {{#5}}
   }
   \spcB \thmNameSize(d) 
   }
 
\newcommand{\thmTALong}[5]{
    \TyHead{#1}{#2}\\
   \thmTLong{#1}{#2}{#3}{#4}{#5}
}

\newcommand{\thmTShort}[5]{
\spcB $\begin{array}{lll}
\sizeof{\ruleSrc{#1}{#2}} &=  {#3} \\
                          &> {#4}  &= \sizeof{\ruleTar{#1}{#2}} 
\end{array}$
}
\newcommand{\thmTAShort}[5]{
    \TyHead{#1}{#2}\\
\thmTShort{#1}{#2}{#3}{#4}{#5}
}

\newcommand{\thmNPros}[3]{
e$_x$={#1} \\
\spcB Show \thmNameN(x)  with proof by contradiction.
Assume C\thmNameN(Q$_x$)\\ 
\spcB case  on structure {#3}\\
}

\newcommand{\normalProse}[2]{
&\text{If }e_{#1}={#2} &
}
\newcommand{\normalLine}[2]{
\normalProse{#1}{#2} \text{then }Q(e_x)\text{ because }e_x\text{ is in normal form.}\\
}
\newcommand{\normalTy}[2]{
\normalProse{#1}{#2} \text{then }Q(e_x)\text{ because }e_x\text{ is not a supported type.}\\
}
\newcommand{\normalNA}[2]{
\normalProse{#1}{#2} \text{and assuming }Q(e) \text{ then }Q(e_x)\\
}
\newcommand{\normalNAS}[2]{
\normalProse{#1}{#2} \text{and assuming }Q(e_3)\text{ and }Q(e_4) \text{ then }Q(e_x)\\
}

\newcommand{\normalRule}[3]{
\normalProse{#1}{#2} \text{then }Q(e_x) \text{ because we can apply rule }{#3}\\
}

\newcommand{\thmL}[2]{
\TyHead{#1}{#2}\\
\spcB Let \text{y}=  $\ruleSrc{#1}{#2}$  and since \text{y} is not in normal form then \donecaseM{#1}{#2}
  }

\usepackage{mathpartir}

\title{Properties of Normalization for a math based intermediate representation}
\author{Charisee Chiw and John Reppy}

\begin{document}

\maketitle
\abstract{
The Normalization transformation 
plays a key \role{} in the compilation of Diderot programs.
The transformations are complicated and it would be easy for a bug
to go undetected. 
To increase our confidence in normalization part of the compiler we provide a formal analysis on the rewriting system.
We proof that the rewrite system is type preserving, value preserving (for tensor-valued expressions), and terminating.
}

\section{Introduction}
The Diderot language is a domain-specific language for scientific visualization and image analysis [3,4].
Algorithms in this domain are used to visually explore data and compute features and properties.
The language supports a high-level model of computation based on continuous tensor fields.
The users rely on a high level of expressivity to implement visualization techniques.

Internally, we represent these computations with a a concise intermediate representation,  called \name{} [1,2].
Inside the compiler, we generate, compose, normalize, and optimize \name{} operators. 
Unfortunately, the IR can quite large, dense, and impossible to read. 
It can be difficult to validate the correctness of computations represented in this IR.

To address the correctness of our work, we provide the following formal analysis.
We define a type system for \name{} operators and provide  evaluation rules.
We show that the rewriting system is type preserving and value preserving for the tensor valued rules. 
We define a size metric on the structure on an \name{} expression. 
The rewriting system always decrease the size of an expression.
 We define a subset of the \name{} expressions to be \fontemph{normal form}. 
We show that termination implies normal form  and that normal form implies termination.
  For any expression  we can apply rewrites until termination, at which point
 we will have reached a normal form expression.

The paper is organized as follows..
We prove that the rewrite system is type preserving  in \secref{properties:type}. 
In \secref{properties:value} we show that for tensor-valued expressions the rewrite system is value preserving.
Lastly, we show that the rewriting system is terminating in \secref{properties:term}.
We present the full proofs in the appendix.
\section{Type Preservation}
\label{properties:type}
%
\begin{figure}
\setcounter{typecnt}{0}
      
\begin{displaymath}
\judgeBase
\infer{
    \begin{array}{cccc}
      \Gamma(T) = \TensorTy{d_1,\,\ldots,\,d_n} \spcC
    |\alpha| = n \spcC  \sigma \vdash \alpha <[d_1,\dots d_n]
      \end{array}
  }{
    \TyJud{\TyEnv{}}{T_{\alpha}}{\genTTyA{\sigma}}
  }
\end{displaymath}%
\begin{displaymath}
  \infer{
      \begin{array}{cccc}
   \Gamma(F) = \FieldTy{d}{d_1,\,\ldots,\,d_n} \spcC
   |\alpha| = n \spcC  \sigma \vdash \alpha <[d_1,\dots d_n] \\
           \end{array}
  }{
   \TyJud{\TyEnv}{F_{\alpha}}{\genFTyA{\sigma}}
  }
\end{displaymath}%

\begin{displaymath}
\judgeConv
  \infer{
     \begin{array}{cccc}
\Gamma(V) = \ImageTy{d}{d_1,\,\ldots,\,d_n} \spcC
    \Gamma(H) = \KernelTy\\
        |\alpha \beta|  = n \spcC
    \sigma \vdash \alpha\beta  <[d_1,\dots d_n]
        \end{array}
  }{
    \TyJud{\TyEnv{}}{V_\alpha \circledast H^\beta}{\genFTyA{\sigma}}
  }  
  \end{displaymath}%
  \begin{displaymath}
\judgeSum
  \infer{
   \begin{array}{cccc}
    i \not\in\dom(\sigma) \spcC
     \sigma'=\sigma[i\mapsto (1,n)]
       \spc\spc
    \TyJud{\TyEnv'}{e}{ \einTyB{\sigma'}}
     \end{array}
  }{
    \TyJud{\TyEnv{}}{\sum\limits^{n}_{i=1} e}{\einTyB{\sigma}
  }}
\end{displaymath}%
\begin{displaymath}
\judgeDeriv
  \infer{
  \begin{array}{cc}
    \sigma(i) = d 
    \spcD 
    \sigma' = \sigma \setminus i   \spcD
    \TyJud{\TyEnv'}{e}{\genFTyA{\sigma'}}
   \end{array} 
  }{
    \TyJud{\TyEnv{}}{\frac{\partial}{\partial x_{i}}  e}{\genFTyA{\sigma}}
  }
\end{displaymath}%
  \begin{displaymath}
\judgeDel
  \begin{array}{ccc}
    \infer{
      i,j \in dom(\sigma)
    }{
      \TyJud{\TyEnv{}}{\delta_{ij}}{\genTTyA{\sigma}}
    }
    & \quad &
       \infer{\TyEnv \vdash ok}
     {\linegamT{ \delta_{\cdot} \delta_{\cdot} }}
  \end{array}
\end{displaymath}%

\begin{displaymath}
\judgeProdDel
  \begin{array}{ccccc}
    \infer{
        \sigma'={\sigma[j\mapsto (1,d)]/\ i} \spcB
      \TyJud{\TyEnv'}{e}{\einTyB{\sigma' }} 
    }{
      \TyJud{\TyEnv{}}{(\delta_{ij} * e)}{\einTyB{\sigma}}
    }
  \end{array}
\end{displaymath}%
  \begin{displaymath}
\judgeProdEps
  \begin{array}{ccc}
    \infer{
     \forall i \in \alpha. i \in dom(\sigma)
    }{
      \TyJud{\TyEnv{}}{\mathcal{E}_{\alpha}}{\genTTyA{\sigma}}
    }
    & \quad & 
    \linegamT{\ruleSrc{A}{1}} 
      \end{array}
\end{displaymath}%
\begin{displaymath}
    \infer{
      \TyJud{\TyEnv}{e}{\tau}
    }{
      \TyJud{\TyEnv}{(\mathcal{E}_{\alpha} * e)}{\tau}
    }
\end{displaymath}%
\caption[Typing Judgments]{Typing Rules for each \name{} expression.}
\label{fig:tyJudge1}
\end{figure}

\begin{figure}
\begin{displaymath}
\judgeProbe
    \infer{
      \begin{array}{cc}
      \TyJud{\TyEnv{}}{\delta_{ij} }{\tau} 
      \\
          \TyJud{\TyEnv{}}{x}{\TensorTy{d}}
      \end{array}
    }{
      \TyJud{\TyEnv{}}{\delta_{ij} @ x}{\tau}
    }
    \spc\spc
    \infer{
        \begin{array}{cc}
      \TyJud{\TyEnv{}}{\mathcal{E}_{\alpha} }{\tau} 
        \\
          \TyJud{\TyEnv{}}{x}{\TensorTy{d}}
     \end{array}
    }{
      \TyJud{\TyEnv{}}{\mathcal{E}_{\alpha} @ x}{\tau}
    }
    \spc\spc
    \infer{
        \begin{array}{cc}
      \TyJud{\TyEnv{}}{e}{\genFTyA{\sigma}} \spc
  \\
          \TyJud{\TyEnv{}}{x}{\TensorTy{d}}
    \end{array}
    }{
      \TyJud{\TyEnv{}}{e @ x}{\genTTyA{\sigma}}
    }
\end{displaymath}%
\begin{displaymath}
\judgeLift
  \begin{array}{ccc}
    \infer{
      \TyJud{\TyEnv{}}{e}{\genTTyA{\sigma}}
    }{
      \TyJud{\TyEnv{}}{\mathbf{lift}_d (e)}{\genFTyA{\sigma}}
    }
  \end{array}
\end{displaymath}%
\begin{displaymath}
\judgeM
    \infer{
      \TyJud{\TyEnv{}}{e}{\einTyB{}} \spc
      \text{$\odot_1\in\SET{\sqrt{}, -, \kappa, \exp, {(\cdot)}^n}$}
    }{
      \TyJud{\TyEnv{}}{\odot_1(e)}{\einTyB{}}
    }
\end{displaymath}%
\begin{displaymath}
\judgeAdd{}
    \infer{
      \TyJud{\TyEnv}{e_1}{\tau} \spc
      \TyJud{\TyEnv}{e_2}{\tau} \spc
      \text{$\odot_2\in\SET{+, -}$}
    }{
      \TyJud{\TyEnv}{(e_1 \odot_2 e_2)}{\tau}
    }
 \spc\spc
    \infer{
      \TyJud{\TyEnv}{e}{\tau}
    }{
      \TyJud{\TyEnv}{{-}e}{\tau}
    }
\end{displaymath}%
\begin{displaymath}
\judgeProd{}
    \infer{
      \TyJud{\TyEnv}{e_1}{\einTyA{}} \spc
      \TyJud{\TyEnv}{e_2}{\einTyA{}}
    }{
      \TyJud{\TyEnv}{(e_1 * e_2)}{\einTyA{}}
    }
\end{displaymath}%
\begin{displaymath}
\judgeDiv{}
    \infer{
      \TyJud{\TyEnv}{e_1}{\einTyB{\sigma}} \spc
      \TyJud{\TyEnv}{e_2}{\einTyB{}}
    }{
      \TyJud{\TyEnv}{\frac{e_1}{e_2}}{\einTyB{\sigma}}
    }
\end{displaymath}%
\caption[Typing Judgments (continued)]{Typing Rules for each \name{} expression.}
\label{fig:tyJudge2}
\end{figure}

\subsection{Typing \name{} Operators}
At the level of the SSA representation, we have types $\theta\in\textsc{Type}$ that correspond to the
surface-level types:\\
$$\begin{array}{llll}
  \theta& :: = &\TensorTy{d_1,\,\ldots,\,d_n}& \text{tensors}\\
 &\mid &\FieldTy{d}{d_1,\,\ldots,\,d_n} &\text{fields}\\
&  \mid &\ImageTy{d}{d_1,\,\ldots,\,d_n}& \text{images}\\
   &  \mid &\KernelTy &\text{kernels}
\end{array}$$
An \name{} operator $\EinOp{\bar{x}}{e}{\sigma}$ can then be given a function type
$(\theta_1 \times \cdots \times \theta_n) \rightarrow \theta$, where $\theta$ is
either $\TensorTy{d_1,\,\ldots,\,d_n}$ or $\FieldTy{d}{d_1,\,\ldots,\,d_n}$
and $\sigma$ is $1<i_1<d_1,\,\ldots,\,1<i_n<d_n$.
The \name{} expression ($e$) is the body of the operator, cannot be given a type $\theta$, however since it
represents a computation indexed by $\sigma$.
Thus the type system for \name{} expressions must track the index space as part of the context.\\
\\
We define the syntax of indexed \name{}-expression types as
\begin{displaymath}
\begin{array}{ll}
  \tau_{0}:: = \genTTyB{} \mid \genFTyB{}\\
  \einTyA :: = \einTyB{\sigma}
  \end{array}
\end{displaymath}%
where $\genTTyA{\sigma}$ is the type of indexed tensors and $\genFTyA{\sigma}$
is the type of indexed $d$-dimensional fields.
We define our typing contexts as
$\TyEnv \in (\textsc{Var} \finmap\tau)^{*} \times (\textsc{IndexVar} \finmap(\mathbb{Z}  \times\mathbb{Z}))^{*} $.
The typing context $\Gamma,\sigma$ includes both the index map and an assignment of types to non-index variables.

With $\Gamma$ we key the map with a variable.
The notation   $$\Gamma(V) = \ImageTy{d}{d_1,\,\ldots,\,d_n}$$
indicates that we can look up  parameter id $V$ in $\Gamma$ and find the resulting type.

We key the map with an index $\sigma\in(\textsc{IndexVar} \finmap(\mathbb{Z}  \times\mathbb{Z})) ^{*}$.
To recall, the notation $i:n$ represents the upper boundary $1<i<n$.
We use notation $$\sigma(i)=n$$ to indicate that we can look up variable ($i$) in $\sigma$ and the upper bound of the variable is n.
It is helpful to view $\sigma$ as defining a finite map from index variables to the size
of their range.
To indicate the addition of a binding  we use ``$\sigma=\sigma'[i\mapsto (1,n)]$".
The domain of $\sigma$ is a sequence, which has to be disjoint
 ($\dom(\sigma)=  \SET{i_1, \dots, i_n}$).
We use  $i \not\in\dom(\sigma)$ to show that $i$ is not in $\sigma$.
We use ``$\sigma=\sigma' \setminus i$" to indicate that i is not in $\sigma'$ but it is in $\sigma$.

We state $\vdash\TyEnv\;\mathbf{ok}$ to show that the environment is okay and the following apply
\begin{itemize}[noitemsep]
\item with $\sigma$ we key the map with an index and index variables do not repeat $\in \dom(\sigma)$.
\item in $\Gamma$ we key the map with a unique variable parameter.
\end{itemize}
We define judgement form $\TyEnv \vdash e:\einTyA{}$ to mean that if the environment is okay then \name{} expression e has type $\einTyA{}$.

 We define the judgement  $\sigma \vdash \alpha <[d_1,\dots d_n]$ as a shorthand for the following judgement.  
 \begin{displaymath}    
\infer{ \forall \mu_i\in\alpha\text{, either } \mu_i \in\mathbb{N}\text{ and }1\leq{}\mu_i\leq{}d_i \text{ or }\sigma(\mu_i) = d_i}
{\sigma \vdash \alpha <[d_1,\dots d_n]}     
\end{displaymath}%
  Recall that an \name{} index $\mu$ is either a constant ($\mu \in \mathbb{N}$)  or a variable index $\mu \in \dom(\sigma)$
  
We present a few typing rules next
and refer the reader to \figref{fig:tyJudge1}  and \figref{fig:tyJudge2}  for a complete list of the rules.
First consider the base case of a tensor variable $T_{\alpha}$; the typing rule is
\begin{displaymath}
\infer{
    \begin{array}{cccc}
      \Gamma,{\sigma}(T_{\alpha}) = \TensorTy{d_1,\,\ldots,\,d_n} \spcC
    |\alpha| = n \spcC \sigma \vdash \alpha <[d_1,\dots d_n]
      \end{array}
  }{
    \TyJud{\Gamma,{\sigma}}{T_{\alpha}}{\genTTyA{\sigma}}
  }
\end{displaymath}%
The antecedents of this rule state that $T_{\alpha}$ has a type that is compatible with
both the multi-index $\alpha$ and the index map $\sigma$.
A similar rule applies for field variables.
The rule for convolution yields an indexed field type.
\begin{displaymath}
  \infer{
     \begin{array}{cccc}
\Gamma(V) = \ImageTy{d}{d_1,\,\ldots,\,d_n} \spcD
    \Gamma(H) = \KernelTy\\
        |\alpha \beta|  = n \spcC
    \sigma \vdash \alpha\beta  <[d_1,\dots d_n]
        \end{array}
  }{
    \TyJud{\Gamma,{\sigma}}{V_\alpha \circledast H^\beta}{\genFTyA{\sigma}}
  }  
  \end{displaymath}%

Note that the index space covers both the shape of the image's range and the differentiation
indices.
Consider the following typing judgement for the \name{} summation form:
\begin{displaymath}
  \infer{
   \begin{array}{cccc}
    i \not\in\dom(\sigma) \spcC
     \sigma'=\sigma[i\mapsto (1,n)]
       \spcC
    \TyJud{\Gamma, \sigma'}{e}{\genTTyA{\sigma'}}
     \end{array}
  }{
    \TyJud{\Gamma,{\sigma}}{\sum\limits^{n}_{i=1} e}{\genTTyA{\sigma}}
  }
\end{displaymath}%
Here we extend the index map with $i:n$ when checking the body of the summation $e$.
This rule reflects the fact that summation contracts the expression.
We use a similar rule for differentiation.
\begin{displaymath}
  \infer{
  \begin{split}
    \sigma(i) = d 
    \spcE
    \sigma' = \sigma \setminus i   \spcC
    \TyJud{\TyEnv'}{e}{\genFTyA{\sigma'}}
   \end{split} 
  }{
    \TyJud{\Gamma,{\sigma}}{\frac{\partial}{\partial x_{i}}  e}{\genFTyA{\sigma}}
  }
\end{displaymath}%
 We can look up index $i$ in $\sigma$ with $\sigma(i)=d$ which indicates $ 1 \leq i \leq d$.  
The term $ \sigma' = \sigma \setminus i$ indicates that the index map $\sigma'$ has all the same index bindings as $\sigma$ except $i$. 

The term $\delta_{ij}$ does not change the context.
  \begin{displaymath}
  \begin{array}{ccc}
    \infer{
      i,j \in \dom(\sigma)
    }{
      \TyJud{\Gamma,{\sigma}}{\delta_{ij}}{\genTTyA{\sigma}}
    }
    & \quad &
       \infer{\TyEnv \vdash ok}
     {\linegamT{ \delta_{\cdot} \delta_{\cdot} }}
  \end{array}
\end{displaymath}%
The application of a Kronecker delta function $\delta_{ij}$ adds index $j$ to the context and removes index $i$. 
\begin{displaymath}
  \begin{array}{ccccc}
    \infer{
        \sigma'={\sigma[j\mapsto (1,d)]/\ i} \spcC
      \TyJud{\TyEnv'}{e}{\einTyB{\sigma' }} 
    }{
      \TyJud{\Gamma,{\sigma}}{(\delta_{ij} * e)}{\einTyB{\sigma}}
    }
  \end{array}
\end{displaymath}%
Similarly, the $\mathcal{E}$ term by itself does not change the context.
  \begin{displaymath}
  \begin{array}{ccc}
    \infer{
     \forall i \in \alpha. \spcB i \in dom(\sigma)
    }{
      \TyJud{\Gamma,{\sigma}}{\mathcal{E}_{\alpha}}{\genTTyA{\sigma}}
    }
    & \quad & 
    \linegamT{\ruleSrc{A}{1}} 
      \end{array}
\end{displaymath}%
When applying $\mathcal{E}$ to another term we preserve that term's type.
  \begin{displaymath}
    \infer{
      \TyJud{\Gamma,{\sigma}}{e}{\tau}
    }{
      \TyJud{\Gamma,{\sigma}}{(\mathcal{E}_{\alpha} * e)}{\tau}
    }
\end{displaymath}%

The Probe operation probes an expression  and a tensor $\TensorTy{d}$.
\begin{displaymath}
    \infer{
      \begin{array}{cc}
      \TyJud{\Gamma,{\sigma}}{\delta_{ij} }{\tau} 
      \\
          \TyJud{\Gamma,{\sigma}}{x}{\TensorTy{d}}
      \end{array}
    }{
      \TyJud{\Gamma,{\sigma}}{\delta_{ij} @ x}{\tau}
    }
    \spcC
    \infer{
        \begin{array}{cc}
      \TyJud{\Gamma,{\sigma}}{\mathcal{E}_{\alpha} }{\tau} 
        \\
          \TyJud{\Gamma,{\sigma}}{x}{\TensorTy{d}}
     \end{array}
    }{
      \TyJud{\Gamma,{\sigma}}{\mathcal{E}_{\alpha} @ x}{\tau}
    }
    \spcC
    \infer{
        \begin{array}{cc}
      \TyJud{\Gamma,{\sigma}}{e}{\genFTyA{\sigma}} \spc
  \\
          \TyJud{\Gamma,{\sigma}}{x}{\TensorTy{d}}
    \end{array}
    }{
      \TyJud{\Gamma,{\sigma}}{e @ x}{\genTTyA{\sigma}}
    }
\end{displaymath}%
Consider lifting a tensor term to the field level:\\
\begin{displaymath}
  \begin{array}{ccc}
    \infer{
      \TyJud{\Gamma,{\sigma}}{e}{\genTTyA{\sigma}}
    }{
      \TyJud{\Gamma,{\sigma}}{\lift{e}}{\genFTyA{\sigma}}
    }
  \end{array}
\end{displaymath}%
The sub-term $e$ has a tensor type $\genTTyA{\sigma}$ but the lifted term $\lift{e}$ has a field type $\genFTyA{\sigma}$.  The rest of the judgements are quite straightforward. 
Some unary operators $\{\sqrt{}, -, \kappa, \exp, {(\cdot)}^n\}$ can only be applied to scalar valued terms such as reals and scalar fields.
\begin{displaymath}
    \infer{
      \TyJud{\Gamma,{\sigma}}{e}{\einTyB{}} \spc
      \text{$\odot_1\in\SET{\sqrt{}, -, \kappa, \exp, {(\cdot)}^n}$}
    }{
      \TyJud{\Gamma,{\sigma}}{\odot_1(e)}{\einTyB{}}
    }
\end{displaymath}%
The subexpressions in an addition or subtraction expression have the same type as the result. 
\begin{displaymath}
\judgeAdd
    \infer{
      \TyJud{\Gamma,{\sigma}}{e_1}{\einTyA{}} \spcC
      \TyJud{\Gamma,{\sigma}}{e_2}{\einTyA{}} \spcC
      \text{$\odot_2\in\SET{+, -}$}
    }
    {
      \TyJud{\Gamma,{\sigma}}{(e_1 \odot_2 e_2)}{\einTyA{}}
    }
 \end{displaymath}%
%
%
%
The full set of typing judgements and corresponding inversion lemmas are contained in \figref{fig:tyJudge1}, \figref{fig:tyJudge2}, and \figref{fig:tyInv}, respectively.
\begin{displaymath}
  \infer{
    \sigma = i_1:d_1,\,\ldots,\,i_m:d_m \\
    \TyJud{(\sigma, \SET{x_i\mapsto\theta_i \mid 1\leq{}i\leq{}n})}{e}{\genTTyA{\sigma}}
  }{
    \TyJud{}{\EinOp{(x_1:\theta_1,\,\ldots,\,x_n:\theta_1)}{e}{\sigma}}{%
      (\theta_1 \times \cdots \times \theta_n) \rightarrow \TensorTy{d_1,\,\ldots,\,d_m}}
  }
\end{displaymath}%

\begin{figure}
\setcounter{typecnt}{0}
\begin{longtable}{lllll}
\noindent
\invgamN{ T_{\alpha}}{\tau}{\tau =\genTTyA{\sigma}}\\
\invgamN{F_{\alpha}}{\tau  }{\tau =\genFTyA{\sigma}}\\
$\dots$
\end{longtable}
\caption[Inversion Lemma for Typing Judgements]{The inversion lemma  makes inferences  based on a structural type judgements. Given a conclusion (left), we can infer something about the type $\tau$ (right).}
\label{fig:tyInv}
\end{figure}

\subsection{Type preservation Theorem}

\setcounter{typecnt}{0}
Given the type system for \name{} expressions presented above, we prove that types are preserved
by normalization.

\begin{theorem}[Type preservation]
  \label{thm:type}
  If $\vdash\TyEnv\;\mathbf{ok}$, $\TyJud{\TyEnv}{e}{\tau}$, and
  $e \rewriteRule{} e'$, then $\TyJud{\TyEnv}{e'}{\tau}$
\end{theorem}%
Given a derivation $d$ of the form $e \rewriteRule{} e'$ we state \thmNameT(d) as a shorthand for the claim that the derivation preserves 
the type of the expression $e$.
For each rewrite rule ($e \rewriteRule{} e'$), the structure of the left-hand-side (LHS) term determines the last typing rule(s) that apply
in the derivation of $\TyJud{\TyEnv}{e}{\tau}$.
We then apply a standard inversion lemma and derive the type of the right-hand-side (RHS) of the rewrite.
Provided below are key cases of the proof (\secref{proof:typePreserv}).
 \begin{description}
 \item[\ruleNum{B}{4}]
The rewrite rule (R4) has the form  $\ruleEin{B}{4}$.\\
The left hand side of the rewrite rule is a tensor type because it is the result of a probe operation.
\TyLHS{B}{4}{\genTTyA{\sigma}}
\spcB { \infer
     {
     \infer{\TyEnv[i\mapsto (1,n)] \vdash e_1:\genFTyA{\sigma[i\mapsto (1,n)]} \invSum}
           {\TyEnv \vdash  (\sum\limits_{i=1}^{n} (e_1)):\genFTyA{\sigma}\invProbe} 
     \spcC 
      \TyJud{\Gamma,{\sigma}}{x}{\TensorTy{d}}}
     {\TyEnv \vdash  (\sum\limits_{i=1}^{n}  (e_1))@x:\genTTyA{\sigma}}}\\
\TyLastFind{}
\TyGiven{\linegam{e_1}{\genFTyA{\sigma[i\mapsto (1,n)]}}}{}\\
\spcB then 
\TyApply{\judgeProbe}{ \linegam{e_1@x}{\genTTyA{\sigma[i\mapsto (1,n)]}}}\\
\spcB and 
\TyApply{\judgeSum}{\linegamT{\ruleTar{B}{4}}}\\
\donecaseT{B}{4}
 \item[\ruleNum{C}{14}]
The rewrite rule (R6) has the form  $\ruleEin{C}{14}$.\\ 
The left hand side of the rewrite rule is a field type because it is the result of a field operation.
\TyLHS{C}{14}{\genFTyA{\sigma}}
We use inversion to find the type for subexpressions $e_1$ and $e_2$.\\
{  \spcB \inferruleLayer
    {\TyEnv \setminus i \vdash e_1 \spcD e_2:\genFTyA{\sigma \setminus i}, 
       \invProd}
    {(\TyEnv \setminus i \vdash e1*e2:\genFTyA{\sigma \setminus i}
        \invDeriv }
    {\TyEnv \vdash\einpart{i:d}(e1*e2):\genFTyA{\sigma} }
}\\
\TyLastFind{}
\TyGiven{\linegam{e_1,e_2}{\genFTyA{\sigma \setminus i}}}{ }\\ 
\spcB
 then \TyApply{\judgeDeriv}{\linegamF{ \einpart{i}(e_1), \einpart{i}(e_2)}},
\\ \spcB 
\TyApply{\judgeProd}{\linegamF{e_1 * \einpart{i}(e_2),  e_2 * \einpart{i}(e_1)}}
, \\ \spcB and
\TyApply{\judgeAdd}{\linegamF{e_1 * \einpart{i}(e_2)+ e_2 * \einpart{i}(e_1)}}.\\
 \donecaseT{C}{14}
\item[\ruleNum{C}{11}]
 The rewrite rule (R7) has the form  $\ruleEin{C}{11}$.
  The left hand side of the rewrite rule is a field type because it is the result of a field operation. 
  \TyLHS{C}{11}{\genFTyA{\sigma}}
We use inversion to find the type for subexpressions $e_1$ and $e_2$.\\
\spcB {   \inferruleLayer
    {\TyEnv \setminus i \vdash e_1:\genFTyA{\sigma \setminus i} \spcD\linegamS{ e_2}, 
       \invDiv
    }
    {(\TyEnv \setminus i \vdash \frac{e1}{e2}:\genFTyA{\sigma \setminus i}
        \invDeriv 
    }
    {\TyEnv \vdash\einpart{i:d}(\frac{e1}{e2}):\genFTyA{\sigma} 
    }}
\\
\newline \TyLastFind{}
We use a type judgement to get the type of the subexpressions ($e_2*e_2$) in the right hand side of the rewrite rule.\\
\TyGiven{\linegamS{e_2}}{}
then \TyApply{\judgeProd}{\linegamS{e_2 * e_2}}\\
We use a type judgement to get the type of the subexpressions ($\einpart{i:d} e_2$) in the right hand side of the rewrite rule.\\
\TyGiven{\linegamS{e_2}}{}
then
\TyApply{\judgeDeriv}{\linegamD{\einpart{i:d}  e_2}}\\
Next, we use a type judgement to get the type of the subexpressions ($ e_1*\einpart{i:d}e_2$) in the right hand side of the rewrite rule.\\
\TyGiven{\linegamD{\einpart{i:d}  e_2 }}{}\\
  \spcB 
\text{ and }$\linegam{e_1}{\genFTyA{\sigma \setminus i}}$
\\  \spcB then
\TyApply{\judgeProd}{\linegamF{e_1 \einpart{i:d}  e_2}}\\
The same is done to find $\linegamF{e_2 \einpart{i:d}  e_1}$\\
\TyGiven{\linegamF{(( \einpart{i}  e_1)* e_2), (e_1 * \einpart{i}  e_2) }}{}
\\ \spcB  \text{and }$\linegamS{e_2*e_2}$
\\ \spcB then 
\TyApply{\judgeAdd}{\linegamF{(( \einpart{i}  e_1)* e_2) -(e_1 * \einpart{i}  e_2)}}
\\ \spcB and
\TyApply{\judgeDiv}{\linegamF{\ruleTar{C}{11}}}\\
 \donecaseT{C}{11}
  \item[\ruleNum{C}{8}] 
 The rewrite rule (\ruleNum{C}{8}) has the form  $\ruleEin{C}{8}$.\\ 
   The left hand side of the rewrite rule is a field type because it is the result of a field operation.
\TyLHS{C}{8}{\genFTyA{i}}  
   We use inversion to find the type for subexpression $e_1$.\\
\spcB {\inferruleLayer
    { \linegamS{e_1} \invM}
    {  \linegamS{ \textbf{sine} (e_1)}}
   { \TyEnv[i\mapsto (1,d)]  \vdash \einpart{i} (\textbf{sine} (e_1)):\genFTyA{i}}
}\\
\\
\TyLastFind{}
\TyGiven{\linegamS{  e_1}}{ }
\\ \spcB  
then
\TyApply{\judgeDeriv}{ \linegamD{\einpart{i}  e_1} },
\\ \spcB 
\TyApply{\judgeM}{ \linegamS{\textbf{cosine}(e_1)} },
\\ \spcB  
and
\TyApply{\judgeProd}{\linegamD{( \textbf{cosine}(e_1))*(\einpart{i}  e_1)} }.\\
\donecaseT{C}{8}

\item[\ruleNum{E}{1}] 
 The rewrite rule (\ruleNum{E}{1}) has the form  $\ruleEin{E}{1}$.\\ 
   We use inversion to find the type for subexpression $e_1,e_2,e_3$.\\
\TyLHS{E}{1}{$\einTyB{\sigma}$}
{ \spcB 
    \inferrule
    {    \inferrule
        {\linegamA{ e_1}, \TyEnv \vdash e_2:\einTyB{} \invDiv}
        {\linegamA{\frac{e_1}{e_2}}}
     \spcD 
     e_3:\einTyB{}\invDiv
    }
    {\linegamA{\frac{\frac{e_1}{e_2}}{e_3}}}
}\\
\TyLastFind{}
\TyGiven{\linegamT{e_1}, \linegamR{e_2,e_3}}{ }
\\ \spcB
then 
\TyApply{\judgeProd}{ \linegamR{ e_2*e_3}}{},
\\ \spcB 
and
\TyApply{\judgeDiv}{ \linegamT{ \frac{e_1 }{ e_2 e_3}}}.\\
\spcB T(\ruleNum{E}{1} \text{ for } $\tau= \genTTyA{\sigma}$) \\
 \donecaseT{E}{1} 
\item[\ruleNum{A}{7}] 
The rewrite rule (\ruleNum{A}{7}) has the form  $\ruleEin{A}{7}$.\\
We define a few variables $\sigma_2 ={\sigma'  /\ ij}$ , $\sigma_j ={\sigma'  j/\ i}$, and  $\sigma_i ={\sigma' i/\ j}$\\ 
We  claim the type for the subexpression ($e_1$).$\TyJud{\Gamma , \sigma_2}{e_1}{\genFTyA{\sigma_2}}$\\
We use a type judgement to get the type of the subexpression ($\einpart{j}  e_1$).\\
\TyGiven{  \TyJud{\Gamma , \sigma_2}{e_1}{\genFTyA{\sigma_2}}}{}
then 
\TyApply{\judgeDeriv }{\TyJud{\Gamma,{\sigma_j}}{\einpart{j}  e_1}{ \genFTyA{\sigma_j}}}\\
We switch the indices when applying the $\delta_\cdot$ \\
\spcB so that
\TyApply{\judgeProdDel }{ \TyJud{\Gamma,{\sigma_i}}{ \delta_{ij} ( \einpart{j}  e_1  )}{ \genFTyA{\sigma_i}} }\\
\TyLastFind{}
 \TyGiven{ \TyJud{\Gamma , \sigma_2}{e_1}{\genFTyA{\sigma_1}}}{}
then
\TyApply{\judgeDeriv }{ \TyJud{\Gamma,{\sigma_i}}{\einpart{i}  e_1}{ \genFTyA{\sigma_i }}} \\
   \donecaseT{A}{7}
\item[\ruleNum{E}{5}]
 The rewrite rule (R41) has the form  $\ruleEin{E}{5}$.\\ 
   We use inversion to find the type for subexpression $s$ and $e$.\\
 \TyLHS{E}{5}{$\einTyB{\sigma}$}  
{ \spcB
\inferruleLayer
    {\begin{array}{ll}
    \TyEnv' \vdash  s:\einTyB{},  \invProd
    \spcD
     \TyEnv' \vdash e_1:\einTyB{\sigma'}\\
      \end{array}
    }
    {
    \sigma' = \sigma[i\mapsto (1,n)]
    \spcD
    \TyEnv' \vdash  s*e_1:\einTyB{\sigma'}
      \invSum}
    {\linegamA{  (\sum\limits_{i=1}^{n} (s*e))}
    }
}\\
\TyLastFind{}
\TyGiven{\linegam{e_1}{\einTyB{\sigma[i\mapsto (1,n)]}} \text{ and }\linegamAS{s}}{-}
\\ \spcB
 then \TyApply{\judgeSum}{\linegamA{ \sum\limits_{i=1}^{n} (e_1)}}
 \\ \spcB
and \TyApply{\judgeProd}{\linegamA{ s*\sum\limits_{i=1}^{n} (e_1)}}.\\
\donecaseT{E}{5}
\end{description}
\newpage
\section{Value Preservation}
\label{properties:value}

\subsection{Math background}\label{background:basics}
In this section, we describe some additional mathematical concepts used by Diderot.
We define some specific operators and their properties.
These concepts are used in the following description about tensor fields and in other parts of the dissertation.

The permutation tensor
or Levi-Civita tensor is represented in \name{} with  $\mathcal{E}_{ij}$ and $\mathcal{E}_{ijk}$ for the 2-d and 3-d case, respectively.
\begin{equation}
  \begin{array}{ccc}
   \mathcal{E}_{ij} =&  \begin{cases}
     +1 & ij \text{ is (0,1)} \\
   -1 & ij\text{ is (1,0)}\\
     0  & \mathrm{ otherwise }
     \end{cases}
  \end{array}
  \text{ and }
    \begin{array}{ccc}
   \mathcal{E}_{ijk} =&  \begin{cases}
     +1 & ijk \text{ is cyclic (0,1,2)} \\
   -1 & ijk \text{ is anti-cyclic (2,1,0)}\\
     0  & \mathrm{ otherwise }
     \end{cases}
  \end{array}
  \label{ein:eps}
\end{equation}%

\noindent The kronecker delta function is  $\delta_{ij}$.
\begin{equation}
  \begin{array}{ccc}
   \delta_{ij} & = & \begin{cases}
      1 & i=j\\ 0  & \mathrm{otherwise}
    \end{cases}
  \end{array}
  \label{ein:dels}
\end{equation}%
The Krnocker delta value has the following property when two deltas share an index:
\begin{equation}   \delta_{ik}  \delta_{kj}=  \delta_{ij} \label{eq:orb8} \end{equation}
and the following when the indices are equal:
\begin{equation}  \delta_{ii}=3 \label{eq:orb9} \end{equation}

\noindent
We reflect on the following properties that hold in an orthonormal basis [5].
Let us define an orthonormal basis $\beta$ with unit basis vectors as {$b_i, b_j,\dots$}.  
 Each basis vector is linearly independent and normalized such that \\
\begin{equation}
\delta_{ij} \spcA=\spcA b_i \cdot b_j \spcA=\spcA\begin{cases} 1 & \text{ i=j }\\ 0 & \text{ otherwise }\end{cases}
\label{eq:orb1}
\end{equation}
Any vector \textbf{u} can be defined by a linear combination of these basis vectors.
$$\textbf{u}=\sum\limits_i u_i b_i $$
A component of a tensor can be expressed in the following way
\begin{equation}
u_j = \textbf{u} \cdot b_j  
 \label{eq:orb5}
\end{equation}

\subsection{Value Definition}
\setcounter{typecnt}{0}
To show that the rewriting system preserves the semantics of the program, we must give
a dynamic semantics to \name{} expressions.
We assume a set of values ($v\in\textsc{Value}$) that include reals, permutation tensor, Kronecker delta functions, and tensors.
Rather than define the meaning of an expression to be a function from indices to values, we
include a mapping $\rho$ from index variables to indices as part of the dynamic environment.
We define a dynamic environment to be
$\Psi, \rho \in (\textsc{IndexVar}\finmap\mathbb{Z}) \times (\textsc{Var}\finmap\textsc{Value})$,
where $\textsc{Value}$ is the domain of computational values (\eg{}, tensors, \etc{}).
We define the meaning of an \name{} expression (for a subset of \name{} expressions) using a big-step semantics $\EvalJud{\Psi, \rho}{e}{v}$,
where v is a value. 
We describe values next and present evaluation rules  \figref{fig:valueJudgeB}.

\begin{figure}[h]
$$\begin{array}{llll}
\textbf{v} &::=      
& \fontit{Real}(n) 
& n \in \mathcal{R} \\
     &\mid
& {\fontit{Tensor}[{p}\cdot {\basen}]}
&\text{index tensor argument p using basis values }b \\
& \mid &E_{\alpha} & \text{Reduces Levi-Civita  tensor}\\
 &\mid &K_{ij} & \text{Reduces Kronecker delta function}\\
\end{array}$$  
\caption[Value definitions for \name]{Value definitions (\textbf{v}) for a subset of \name{} expression}
\label{fig:values}
\end{figure}  
We assume an orthonormal basis function.
Inspired by \eqqref{eq:orb5}, we use $b_i$ to represent a basis vector inside a value expression.
The value of a vector is defined as 
$$\EvalJud{\Psi, \rho}{T_i}{ \valTensor{T}{b_i}} \\$$
A term $b_i$ is created for each variable index $i$ in the \name{} expressions. 
The full tensor judgement 
$$\EvalJud{\Psi, \rho}{T_{\alpha}}{\valTensor{T}{b_{\alpha1} \dots   b_{\alpha n}  }}$$ is used to represent an arbitrary sized tensor.
The lift operation is used to lift a tensor to a field. 
The value of a lifted term is the value of that term.
$$ \infer{\EvalJud{\Psi, \rho}{e }{ v}  }{ \EvalJud{\Psi, \rho}{\lift{e}  }{v}  }$$


We support  arithmetic operations on and between $u$. 
The summation expression can be evaluated with the following judgement:
$$\infer{ \EvalJud{\Psi, \rho}{e }{v }}{\EvalJud{\Psi, \rho}{\sum\limits_{i=1}^n e}{ \Sigma_{i=1}^n v}}$$
The summation operator is applied to the $u$. Generally, the judgement for unary operators ($\odot_1\in\SET{\Sigma \mid \sqrt{} \mid -\mid \kappa \mid \exp\mid {(\cdot)}^n} $) is as follows:
$$\infer
    { \EvalJud{\Psi, \rho}{e_1 }{ \valReal{r1}}}
    { \EvalJud{\Psi, \rho}{ \odot_1 e_1 }{  \valReal{\odot_1 r1}}}
$$
$$ \infer
    { \EvalJud{\Psi, \rho}{e_1 }{ \valTensor{e_1}{b1}}}
    { \EvalJud{\Psi, \rho}{\odot_1 e_1}{\valopU{\odot_1}{\valTensor{e_1}{b1}} }}$$
The binary operators ($ \odot_2 =+\mid - \mid * \mid /\  $) can be applied between $u$.
.\\
$$
\infer{\EvalJud{\Psi, \rho}{e_1}{\valReal{r_1}} \spc\EvalJud{\Psi, \rho}{ e_2  }{\valReal{r_2}}}{  \EvalJud{\Psi, \rho}{(e_1 \odot_2 e_2) }{   \valReal{r_1  \odot_2 r_2}} }$$
$$ \infer
    {  \EvalJud{\Psi, \rho}{e_1}{\valTensor{e_1}{b1}} \phantom{+++} \EvalJud{\Psi, \rho}{e_2  }{  \valTensor{e_2}{b2}}}
    { 
        \EvalJud{\Psi, \rho}
                {(e_1 \odot_2 e_2) }
                {\valopB{\odot_2}{\valTensor{e_1}{b1}}{\valTensor{e_2}{b2}}}
    } $$
The epsilon and Kronecker delta functions are each reduced to a distinct permutation value ($\valE{\alpha}$ or $\valP{ij}$). 

  $$\EvalJud{\Psi, \rho}{ \mathcal{E}_{ijk}}{\valE{ijk}} \spc \spc \EvalJud{\Psi, \rho}{\delta_{ij} }{\valP{ij}}$$
The value for $\mathcal{E}_{ijk}$ is subject to \eqqref{ein:eps}.
The value for $\delta_{ij} $ is subject to \eqqref{ein:dels}, \eqqref{eq:orb8}, and \eqqref{eq:orb9}.

We use notation $v_1 \mapsto v_2$ to indicate a value that is reduced or rewritten.
 We combine permutation values with tensor values as 
\begin{equation}
 \valP{ij}* \valTensor{T}{\beta} 
\mapsto  \valTensor{T}{b_i\cdot b_j \cdot \beta}.
 \label{eq:valKT}
 \end{equation}
 \noindent The full set of evaluation rules are given in \figref{fig:valueJudgeB}.
 
  \begin{figure}
\begin{longtable}{lllll}
\noindent
$\valJudgmentName{1}$&
$\EvalJud{\Psi, \rho}{c}{ \valReal{c}}$  \\
\\
$\valbaseT $&
$ \EvalJud{\Psi, \rho}{T_{\alpha}}{\valTensor{T}{ b_{\alpha1}  \dots   b_{\alpha n}  }}$\\
\\
$\vallift$
& $\infer{\EvalJud{\Psi, \rho}{e }{ v}  }{\EvalJud{\Psi, \rho}{ \lift{e}  }{ v}  }$\\
\\

$\valunary$
& \text{$\odot_1\in\SET{\sum\limits \mid \sqrt{} \mid -\mid \kappa \mid \exp\mid {(\cdot)}^n} $}\\
& $\infer{ \EvalJud{\Psi, \rho}{e_1 }{\valReal{r1}}}{
\EvalJud{\Psi, \rho}{\odot_1 e_1 }{  \valReal{\odot_1 r1}}}$
\spc $\infer{\EvalJud{\Psi, \rho}{e_1 }{ \valTensor{e_1}{b1}}}{
\EvalJud{\Psi, \rho}{\odot_1 e_1}{ \odot_1  \valTensor{e_1}{b1} }}$\\
\\
$\valbinary$
& $ \odot_2 =+\mid - \mid * \mid /\  $\\
& $\infer{\EvalJud{\Psi, \rho}{e_1 }{ \valReal{r1} }\phantom{+++} \EvalJud{\Psi, \rho}{e_2}{\valReal{r2}}}{\EvalJud{\Psi, \rho}{(e_1 \odot_2 e_2) }{   \valReal{r1  \odot_2 r2}}}$ \\
&\spc $\infer{\EvalJud{\Psi, \rho}{e_1  }{ \valTensor{e_1}{b1}} \phantom{+++} \EvalJud{\Psi, \rho}{e_2 }{  \valTensor{e_2}{b2}}}{\EvalJud{\Psi, \rho}{(e_1 \odot_2 e_2)}{ \valTensor{e_1}{b1} \odot_2 \valTensor{e_2}{b2}}}$ \\
\\
\\
& \valProbeBinRule{}\\
\\
$\valprobe$
&
$\infer{ \EvalJud{\Psi, \rho}{e}{ v}  }{\EvalJud{\Psi, \rho}{\lift{e}@{e} }{ v}}$
\spc $\infer{\EvalJud{\Psi, \rho}{\delta_{ij} }{ v}}{
\EvalJud{\Psi, \rho}{\delta_{ij}@{e} }{ v}}$
\spc
$\infer{
\EvalJud{\Psi, \rho}{\mathcal{E}_{\alpha} }{ v}}{
\EvalJud{\Psi, \rho}{\mathcal{E}_{\alpha}@{e} }{ v}}$\\
\\
$\valeps$&
$ \EvalJud{\Psi, \rho}{\delta_{ij} }{ \valP{ij}}$ \spc 
$\EvalJud{\Psi, \rho}{ \mathcal{E}_{\alpha}}{ \valE{\alpha}}$\\
\end{longtable}
\caption[Value Judgements ]{Value Judgements for each \name{} expression.}
\label{fig:valueJudgeB}
\end{figure}

 \subsection{Value Preservation Theorem}
Our correctness theorem states the rewrite rules  do not change the value of
an expression with respect to a dynamic environment, assuming that the expression and
dynamic environment are both type-able in the same static environment and their value is defined.
\begin{theorem}[Value Preservation]
  \label{thm:value}
  If $\OkJud{}{\Gamma, \sigma}$, $\TyJud{\Gamma, \sigma}{e}{\tau}$,
  $\OkJud{\Gamma, \sigma}{\Psi, \rho}$, $e \rewriteRule{} e'$,
  and $\EvalJud{\Psi, \rho}{e}{v}$,
  then $\EvalJud{\Psi, \rho}{e'}{v}$
\end{theorem}%
Assume $\EvalJud{\Psi, \rho}{e}{v}$ and $e\rewriteRule{} e'$, then the proof proceeds by case analysis of the rewrite rules. 
Does not include rules that involve fields terms (values for fields are not defined).
We show the full proof in \secref{proof:valuePreserv} and select a few key examples below.
 \begin{description}
\item{\ruleNum{D}{3}}
 The rewrite rule (\ruleNum{D}{3}) has the form  $\ruleEin{D}{3}$.\\
$\valHeadingR{D}{3}$
\\ \spc\spc 
then 
\VApply{\valbaseC,\valbinary}{e_1-0  \Downarrow v' -\valReal{0}}.
\\ \spc
The value of $v $ is  $v' -\valReal{0}$.
\\ \spc\spc  By using algebraic reasoning: $v' -\valReal{0} = v'$.
\\ \spc\spc 
Since $e_1-0  \Downarrow v \text{ and }e_1-0  \Downarrow v' \text{ then }  v=v'$\\
$\valEndProse{D}{3}$
\item{\ruleNum{E}{6}}
 The rewrite rule (\ruleNum{E}{6}) has the form  $\ruleEin{E}{6}$.\\
 $\valHeadingR{E}{6}$ 
\\ \spc\spc 
then 
\VApply{\valunary}{\sqrt{e_1} \Downarrow  \valopU{\sqrt}{v'}},
\\ \spc\spc 
and
\VApply{\valbinary}{\sqrt{e_1}\sqrt{e_1} \Downarrow \sqrt{v'}\sqrt{v'}}\\
\spc
The value of $v $ is $ \sqrt{v'}*\sqrt{v'} $\\
\spc By using algebraic reasoning to analyze $v$\\
\spc\spc
\TyBy{\text{reduction}}{v= \sqrt{v'}*\sqrt{v'}}{ v'}\\
$\valEndProse{E}{6}$
\item{\ruleNum{A}{1}}
 The rewrite rule (\ruleNum{A}{1}) has the form  $\ruleEin{A}{1}$.\\
$\valHeading{A}{1}$ 
\spc Given that 
$\mathcal{E}_{ijk} \Downarrow  \valE{ijk}$ 
and $ \mathcal{E}_{pqr} \Downarrow  \valE{pqr}$
 then  
$\mathcal{E}_{ijk} \mathcal{E}_{pqr} \Downarrow \valE{ijk} \valE{pqr}$.\\
\spc The value of $v$ is $ \valE{ijk} \valE{pqr}  $.\\
\spc\spc Consider the product of two $\valE{}$ expressions as \\
\spc\spc  $\valE{ijk} \valE{pqr}
\begin{array}{cccc}\longrightarrow   \left| \begin{array}{ccc} 
\valP{ip} & \valP{iq} & \valP{ir}\\
\valP{jp} & \valP{jq} & \valP{jr}\\
\valP{kp} & \valP{kq} & \valP{kr}
\end{array} \right|\\
 \end{array}$\\
\spc\spc $\longrightarrow  \valP{ip}( \valP{jq}  \valP{kr}-  \valP{jr}  \valP{kq} )+ \valP{iq}( \valP{jr}  \valP{kp}-  \valP{jp}  \valP{kr})+ \valP{ir}( \valP{jp}  \valP{kq}-  \valP{jq}  \valP{kp})$\\ 
\spc \spc  Rewriting so that there is a shared index ($p=i$): \\
\spc\spc  $\longrightarrow 
 \valP{ii}  \valP{jq}  \valP{kr}-  \valP{ii}  \valP{jr}  \valP{kq} +\valP{iq} \valP{jr}  \valP{ki}-   \valP{iq} \valP{ji}  \valP{kr} + \valP{ir} \valP{ji} \valP{kq}-  \valP{ir} \valP{jq}  \valP{ki}$\\ 
\spc\spc Applying  \eqqref{eq:orb9}:\\
\spc\spc  $\longrightarrow 3  \valP{jq}  \valP{kr}-  3  \valP{jr}  \valP{kq} +\valP{iq} \valP{jr}  \valP{ki}-   \valP{iq} \valP{ji}  \valP{kr} + \valP{ir} \valP{ji} \valP{kq}-  \valP{ir} \valP{jq}  \valP{ki} $\\ 
\spc\spc Applying \eqqref{eq:orb8}:\\
\spc\spc  $\longrightarrow 3  \valP{jq}  \valP{kr}-  3  \valP{jr}  \valP{kq} + \valP{kq} \valP{jr}  -   \valP{jq} \valP{kr}  +  \valP{jr}   \valP{kq}-   \valP{kr} \valP{jq} $\\ 
\spc\spc  Reduces to: \\
\spc\spc  $ \longrightarrow \valP{jq}  \valP{kr}-   \valP{jr}  \valP{kq} $\\
\spc\spc Match indices to rule ($q \longrightarrow  l$ and $r \longrightarrow  m$)\\
\spc\spc  $ \longrightarrow \valP{jl}  \valP{km}-   \valP{jm}  \valP{kl} $\\
$\valTarM{A}{1}$
\spc\spc Given that
\VApply{\valdel}{  \delta_{jl}  \Downarrow  \valP{jl}
    \spc \delta_{km} \Downarrow \valP{km}
    \spc  \delta_{jm}   \Downarrow \valP{jm}  
    \spc \delta_{kl}   \Downarrow   \valP{kl}}\\
\spc\spc then
\VApply{ \valbinary}{\delta_{jl}  \delta_{km} \Downarrow \valP{jl}  \valP{km}
        \spc\spc  \delta_{jm}  \delta_{kl}   \Downarrow  \valP{jm}  \valP{kl}}\\
\spc \spc and 
\VApply{ \valbinary}{ \ruleTar{A}{1}\Downarrow \valP{jl} \valP{km}- \valP{jm} \valP{kl}}\\
$\valEndProse{A}{1}$
\item{\ruleNum{A}{5}}
 The rewrite rule (\ruleNum{A}{5}) has the form  $\ruleEin{A}{5}$.\\
 $\valHeading{A}{5}$  
\spc
Given that 
\VApply{\valbaseT{}}{T_j \Downarrow\valTensor{T}{b_j}} 
\\\spc\spc
and \VApply{\valdel{}}{\delta_{ij}  \Downarrow  \valP{ij}}\\
\spc\spc
then 
\VApply{\eqqref{eq:valKT}}{\ruleSrc{A}{5} \Downarrow \valTensor{T}{b_j\cdot b_i\cdot b_j}}\\
\spc
The value of $v $ is $   \valTensor{T}{b_j \cdot b_i\cdot b_j} $\\
\spc By using algebraic reasoning to analyze $v$
\\ \spc\spc 
\TyBy{\text{ reducing value } b_j\cdot b_j \text{ using } \eqqref{eq:orb1}}
{v}{\valTensor{T}{b_i}}\\
$\valTarM{A}{5}$
\spc Lastly, \VApply{\valbaseT{}}{ T_i \Downarrow   \valTensor{T}{b_i}}\\  
$\valEndProse{A}{5}$
 \end{description}
\newpage
\section{Termination}
\label{properties:term}
In this section we make 
the following claims:
\begin{enumerate}
\item Rewriting terminates
\item if e $\rewriteNorm{}$ e' and $\not \exists$ e'' such that e' $\rewriteRule{}$ e'', then e' $\in \mathcal{N}$
\end{enumerate}
We prove that the normalization rewriting
will terminate and that the resulting term will be in normal form.

Our approach uses the standard technique of defining a well-founded 
size metric $\sizee$  to show that
the rewrite rules always decrease the size of an expression.
The size metric guarantees that the normalization
process terminates (\secref{term:size}).
We also want to guarantee that normalization actually produces a normal-form. 
We define a subset of the \name{} expressions that are in \fontemph{normal form} by a grammar \secref{sec:normalform}.
We then define the \fontemph{terminal} expressions as
$\mathcal{T} = \SET{e \mid \text{$\not \exists e'$ such that $e \rewriteRule{}  e'$} }$.
The last section (\secref{term:termnormal}) relates normal form expressions and terminal expressions.
We show that termination implies normal form (\lemref{lem2}) and that normal form implies termination (\lemref{lem3}).
 For any expression  we can apply rewrites until termination, at which point
  we will have reached a normal form expression  (\thmref{proofs:normalization}).

\begin{table}[h]
\caption[Size metric]{We define a size metric $\sizeof{\bullet} : e \rightarrow \mathbb{N}$ inductively
  on the structure of the grammar of EIN in [2].}
\begin{center}
  \begin{tabular}{ll}
    \name{}  expression ($e$)&Size metric  $\LDB\sizee{}$ \\
    \hline
    $c$, $T_\alpha$, $F_\alpha$, $(v_\beta \circledast h^\mu)$, $\delta_{ij}$
					& 1 \\
    $\mathcal{E}_\alpha  $		& 4 \\
    $\lift{e}$, $\sqrt{e}$, ${-}e$, $\exp(e)$, $e^n$, $\kappa(e)$
					& $1+ \sizee$ \\
    $e_1 + e_2$ ,$e_1 - e_2$, $e_1*e_2$ & $1+ \sizea + \sizeb$ \\
    $\frac{a}{b}$			& $2+ \sizea +\sizeb$\\
    $\sum\limits e$				& $2+2 \sizee$\\
    $\einpart{\nu} \diamond e $		& $5^{\sizee} \sizee$\\
    $e(x)$				& $ 2\sizee $ \\
  \end{tabular}
\end{center}%
\label{app:size}
\end{table}
\subsection{Size Metric}
\label{term:size}
We define a size metric $\sizee$ for \name{} expressions in \tblref{app:size} and use it to  show that rewrites always decrease the size of the \name{} expression.

\begin{lemma}
  \label{lem:size}
   If $e \rewriteRule{} e'$ then $\sizee > \sizeof{e'}$ 
\end{lemma}
Our proof does a case analysis on the rewrite rules ($e \rewriteRule{} e'$) and compares the  size (\tblref{app:size}) of each side of the rule.
Provided below are key cases of the proof (\secref{proof:termination1}).
\begin{description}
\item{\ruleNum{B}{1}}
The rewrite rule (\ruleNum{B}{1}) has the form  $\ruleEin{B}{1}$.\\ 
   \spcB case analysis on the operator $\odot_n$ \\
 \spcC if $\odot_n=*$\\
\spcD \thmTComp
    {(e_1 *e_2)@x}
    {2+2\sizea+2\sizeb}
    {1+2\sizea +2\size}
    {\sizeof{(e_1@x) *(e_2@x)}}
 \spcC if $\odot_n=\frac{\bullet}{\bullet}$\\
\spcD \thmTComp
    {\sizeof{(\frac{e_1}{e_2})@x}} 
    {4+2\sizea+2\sizeb}
    {2+2\sizea +2\sizeb}
    {\sizeof{\frac{e_1@x}{e_2@x}}}
 \spcB P(d)
\item{\ruleNum{C}{7}}
The rewrite rule (\ruleNum{C}{7}) has the form  $\ruleEin{C}{7}$.\\
\thmTLong{C}{7}
    {(1+\sizea)5^{(1+\sizea)}}
    {\sizea *(1+5^{\sizea}) +3 }
    {(Lm ~\ref{lemz})}
    \item{\ruleNum{C}{16}}
 The rewrite rule (\ruleNum{C}{16}) has the form  $\ruleEin{C}{16}$.\\
 \thmTLong{C}{16}
{(1+\sizea +\sizeb) 5^{(1+\sizea+\sizeb)}}
{\sizea 5^{(\sizea)}+\sizeb 5^{(\sizeb)}+1}
{(Lm ~\ref{lemy})}
\item{\ruleNum{E}{1}}
  The rewrite rule (\ruleNum{E}{1}) has the form  $\ruleEin{E}{1}$.\\
\thmTLong{E}{1}
    {4+\sizea+\sizeb+\sizeof{e_3}}
    {3+\sizeof{e_1}+\sizeb+\sizeof{e_3}}{}
\end{description}

\subsection{Normal Form}
\label{sec:normalform}
An \name{} expression is in normal form if it can not be reduced.
The normal form is defined as the subset  $\mathcal{N}$ of \name{} expressions.
In the following, we describe the normal form with the following examples.
Some tensors, constants, and permutation terms that are in normal form include:
$$T_\alpha, c \not= 0 , \delta_{ij}, \mathcal{E}_{ij}, \text{ and } \mathcal{E}_{ijk}$$
The field forms $\mathcal{F}$   include:
$$ F_\alpha ,  V \circledast H ,\einpart{i} F_\alpha$$     
All differentiation is applied (via product rule or otherwise) so in normal form the differentiation is only applied to a field term:
$$\einpart{i} F_\alpha $$
until it is pushed down to the convolution kernel:
$$ V \circledast \einpart{i} H$$
The only probed terms are field forms $\mathcal{F}$:
$$F_\alpha @T, (V \circledast H) @x, \text{ and }(\einpart{i}   F) @x$$
Some unary operations are in normal form, as long as their sub-term $e_1$ is in normal form:
$$\text{sine}(e_1), \lift{e_1}, \sqrt{e_1}, \exp({e_1})$$
Other arithmetic operations cannot have a zero constant sub-term [2]
$$-e_1, e_1+e_2,e_1-e_2,e_1*e_2,\frac{e_1}{e_2}$$
The division structure is subject to algebraic rewrites [2]. 
The normal form of the product and summation structure is more restricted in part because of index-based rewrites. 
Normal form is presented more formally next:

 \begin{definition}[Normal Form]
The following grammar specifies the subset $\mathcal{N}$ of \name{} expressions that
are in \fontemph{normal form}:

\begin{displaymath}
  \begin{array}{ccll}
    \mathcal{N} & ::= & \mathcal{ A} \mid c\\
    \mathcal{A} & ::= & \mathcal{D} \mid \mathcal{G}  \\
    \mathcal{D} & ::= & \mathcal{B} \mid -\mathcal{G}  \\
    \mathcal{G} & ::= & \mathcal{B} \mid \frac{  \mathcal{D}}{  \mathcal{D}} \\
    \mathcal{B} & ::= & T_\alpha  \mid \mathcal{F} \mid \mathcal{F} @ T_\alpha
        \mid c \not= 0 \mid \delta_{ij} \mid \mathcal{E}_{ij} \mid \mathcal{E}_{ijk} \\
                & \mid & \mathcal{A}+  \mathcal{A} \mid   \mathcal{A}-\mathcal{A}
         \mid \sqrt{\mathcal{N}}  \\
                & \mid & \lift{\mathcal{N}}\mid \text{exp}(\mathcal{N})
        \mid \mathcal{N}^{c} \mid \kappa(\mathcal{N}) \\
       & \mid &(\mathcal{A}*\mathcal{A})^{1,2,3,4}\\
       & \mid & (\sum \mathcal{N})^{5}\\ 
    \mathcal{F} & ::= & F_\alpha \mid   v \circledast h  \mid  \einpart{i} F_\alpha \\
 \end{array}
\end{displaymath}
subject to the following additional restrictions (noted in the syntax with an upper index):
\begin{enumerate}
  \item If a term has the form $\mathcal{E}_{ijk} * \mathcal{E}_{i'j'k'}$ then
    the indices $ijk$ must be disjoint from $i'j'k'$. 
  \item If a term contains the form $\mathcal{E}_{ijk} * \mathcal{A}$
  and $\mathcal{A}$ has a differentiation  component  then no two of the indices $i,j,$ and $k$ may occur in the differentiation component of $\mathcal{A}$.
  For example, 
  $\mathcal{E}_{ijk}*\einpart{jk} e$ is not in normal form and can be rewritten as $\mathcal{E}_{ijk}*\einpart{jk} e \rewriteNorm{} 0$.
  \item If a term has the form $\delta_{ij} * \mathcal{A} $ then $j$ may not occur in $\mathcal{A}$. 
  For example, the 
  expression $\delta_{ij}* T_j$ is not in normal form, and thus $\delta_{ij}* T_j$ can be rewritten to $T_i$.
    \item If a term has the form $\sqrt{e_1} * \sqrt{e_2}$ then $e_1 \not = e_2$.
   \item If a term is of the form $\sum(e_1 *e_2)$ then $e_1$ can not be a scalar $s$, scalar field $\varphi$, or constant $c$. 
   For example, terms $\sum(s *e_2)$  or  $\sum(\varphi *e_2)$ are not in normal form and can be rewritten as  $s\sum e_2 $ and $\varphi\sum e_2$, respectively.
\end{enumerate}%
\label{app:normal-form}
\end{definition}

\subsection{Termination and Normal form}
\label{term:termnormal}
The following two lemmas relate the set of normal forms expressions to the terminal expressions.
The first shows that termination implies normal form.
\begin{lemma}
  \label{lem2}
  If $e \in \mathcal{T}$, then $e \in \mathcal{N}$ 
\end{lemma}%
  The proof is by examination of the EIN syntax in [2].
  For any syntactic construct, we show that either the term is in normal form, or there
  is a rewrite rule that applies.
  We define $Q(e_x)$ $\equiv \not \exists e_x'$ such that $e_x\rewriteRule{} e_x'$ and $e_x \in \mathcal{N}$. 
The following is a sample of a proof by contradiction (full proof is available \secref{proof:termination2}). \\
\noindent case  on structure $e_x$ \\
$\begin{array}{llll}
\normalLine{x}{c}
\normalLine{x}{T_\alpha}
\normalLine{x}{F_\alpha}
\normalLine{x}{V_\alpha \circledast H}
\normalLine{x}{\delta_{ij}}
\normalLine{x}{\mathcal{E}_\alpha}
&\text{If }e_x=\lift{e_1}  \\
\end{array}$\\
\spcB Prove \thmNameN(e) by contradiction.\\
$\begin{array}{lll}
\spcC
\normalLine{1}{c}
\normalLine{1}{T_\alpha}
\normalTy{1}{F_\alpha}
\normalTy{1}{e \circledast e}
\normalLine{1}{\delta_{ij}}
\normalLine{1}{\mathcal{E}_\alpha}
\normalTy{1}{\lift{e}}
\normalNA{1}{M(e_1)}
& & \text{Given }M(e) = \sqrt{e} \mid exp(e)\mid e_1^n \mid \kappa(e) \\
\normalNA{1}{-e}
\normalTy{1}{\einpart{\alpha}e}
\normalNA{1}{\sum e}
\normalNAS{1}{e_3+e_4}
\normalNAS{1}{e_3-e_4}
\normalNAS{1}{e_3*e_4}
\normalNAS{}{\frac{e_3}{e_4}}
\normalNAS{1}{e_3@e_4}
& Q(e_x)
\end{array}$ \\
The next lemma demonstrates that normal form implies termination.
\begin{lemma}
  \label{lem3}
  If $e \in \mathcal{N}$, then $e \in \mathcal{T}$
\end{lemma}
We state \thmNameL(e) as a shorthand for the claim that if $e$ is in normal form then it has terminated.
The following is a proof by contradiction. 
C\thmNameL(e): There exists an expression e that has not terminated  
and is in normal form.
More precisely, given a derivation $d$ of the form $e \rewriteRule{}e'$ ,
there exists an expression that is the source term $e$ of derivation  $d$ therefore not-terminated, and is in normal form. 
Below are  cases of the proof (\secref{proof:termination3}).\thmL{B}{1}\thmL{B}{2}

\begin{theorem}[Normalization]
  For any closed \name{} expression $e$ the following two properties hold:
  \begin{enumerate}
    \item there exists an \name{} expression $e' \in \mathcal{N}$, such that $e \rewriteNorm{} e'$, and
    \item there is no infinite sequence of rewrites starting with $e$.
  \end{enumerate}%
  In other words, for any expression $e$ we can apply rewrites until termination, at which point
  we will have reached a normal form expression $e'$.
  \label{proofs:normalization}
\end{theorem}

  The theorem follows from Lemmas~\ref{lem:size}, \ref{lem2}, and~\ref{lem3} described in \secref{proof:termination}.

\section{Discussion}
The properties that we have described demonstrate the correctness of the
normalization transformations for \name.
Unfortunately, the rewriting system is not confluent (because different pairings of $\mathcal{E}_{ijk}$
can be rewritten and produce different normal forms). 
In our system, we
apply rules in a standard order, but there may be opportunities for improving performance
by tuning the order of rewrites.

While there are still many opportunities for compiler bugs, normalization is the most critical
part of compiling tensor-field expressions down to executable code, so these results increase 
our confidence in the correctness of the compiler.
There are other parts of the compiler pipeline for which we hope to prove correctness
in the future.

\section*{Bibliography}
\begin{enumerate}
\item
 Charisee Chiw. Ein notation in diderot. Master’s thesis, University of Chicago, April 2014.
\item Charisee Chiw, Gordon L Kindlman, and John Reppy. EIN: An intermediate representation for compiling tensor calculus. In Proceedings of the 19th Workshop on Compilers for Parallel Computing (CPC 2019), July 2016.
\item Charisee Chiw, Gordon Kindlmann, John Reppy, Lamont Samuels, and Nick Seltzer. Diderot: A parallel dsl for image analysis and visualization. SIGPLAN Not., 47(6):111–120, June 2012.
\item Gordon Kindlmann, Charisee Chiw, Nicholas Seltzer, Lamont Samuels, and John Reppy. Diderot: a domain-specific language for portable parallel scientific visualization and image analysis. IEEE Transactions on Visualization and Computer Graphics (Proceedings VIS 2015), 22(1):867–876, January 2016
 \item Gerhard A. Holzapfel. Nonlinear Solid Mechanics. John Wiley and Sons, West Sussex, England, 2000.
 \end{enumerate}

\appendix

\section{Type Preservation Proof}

\label{proof:typePreserv}
The following is a proof for Theorem \ref{thm:type}

Given a derivation $d$ of the form $e  \rewriteRule{} e'$ we state \thmNameT(d) as a shorthand for the claim that the derivation preserves 
the type of the expression $e$.
For each rule, the structure of the left-hand-side term determines the last typing rule(s) that apply
in the derivation of $\TyJud{\TyEnv}{e}{\tau}$.
We then apply a standard inversion lemma and derive the type of the right-hand-side of the rewrite .
The proof demonstrates that $\forall d. T(d)$.
\newline
\noindent Case on structure of d
\TyHead{B}{1}\\
\TybodyCase{ $\odot_n$ }{$*$}{$/$}
if $\odot_n=*$\\ 
\TyFind{(e_1 *e_2)@x}{}
\TybodyProbe
\TyLHS{B}{1}{\genTTyA{\sigma}}\\
{\spcC
    \inferrule
        {\inferrule
            { \linegamF{  e_1}\spcC \linegamF{ e_2}\invProd}
            { \linegamF{e1*e2} \invProbe}
         \spcD \TyJud{\Gamma,{\sigma}}{x}{\TensorTy{d}}
        }
    { \linegamT{(e1*e2)@x} 
    }
}\\
\TyLastFind{}
\TyFind{(e_1@x) *(e_2@x)}
\TyGiven{\linegamF{e_1,e_2}}{},
\\ \spcB then
\TyApply{\judgeProbe}{ \linegamT{e_1@x,e_2@x}}, 
\\ \spcB and
\TyApply{\judgeProd }{ \linegamT{ e_1@x* e_2@x}}\\
\spcB T(\ruleNum{B}{1} \text{ for } $\odot_n=*$) \\
if $\odot_n=/$\\
\TyFind{(\frac{e_1}{e_2})@x}
\TybodyProbe
\spcB \TybodySrcBB{B}{1}{\genTTyA{\sigma}}{\invProbe }\\
\TyFind{e_1 \text{ and } e_2}
{\spcC 
\inferrule
    {\inferrule
        {\linegamF{e_1},\spcD \TyEnv \vdash e_2:\genFTyA{} \invDiv}
        {\TyEnv \vdash  (\frac{e1}{e2}):\genFTyA{\sigma}\invProbe}
      \spcD \TyJud{\Gamma,{\sigma}}{x}{\TensorTy{d}}
     }
    {\TyEnv \vdash (\frac{e1}{e2})@x:\genTTyA{\sigma}}}\\
\TyFind{\frac{(e_1@x)}{(e_2@x)}}
\TyGiven{\linegamF{e_1},\linegamS{e_2}}{
}
\\\spcB
then 
\TyApply{\judgeProbe}{ \linegamT{e_1@x}}, 
\\ \spcB
\TyApply{\judgeProbe}{ \linegamR{e_2@x}},
\\ \spcB
and \TyApply{\judgeDiv }{ \linegamT{\frac{e_1@x}{ e_2@x}}}.  \\
  \spcB T(\ruleNum{B}{1} \text{ for } $\odot_n=/$)\\
\donecaseT{B}{1}
\TyHead{B}{2} \\
\spcB   $\odot_2=+ \mid -$\\
\TyFind { (e_1 \odot_2 e_2)@x}
\TybodyProbe
\TyLHS{B}{2}{\genTTyA{\sigma}}
\spcC
\inferrule
    {\inferrule
        {\linegamF{ e_1,e_2} \invAdd}
        {\linegamF{e_1 \odot_2 e2}\invProbe}
     \spcD \TyJud{\Gamma,{\sigma}}{x}{\TensorTy{d}}
     }
    {\linegamT{(e1 \odot_2 e2)@x}}\\
\TyLastFind{}
\TyGiven{\linegamF{e_1,e_2}}{}
\\ \spcB then 
\TyApply{\judgeProbe}{ \linegamT{e_1@x, e_2@x}}
\\\spcB and
\TyApply{\judgeAdd}{\linegamT{e_1@x \odot_2 e_2@x}}
\TyHead{B}{3}\\
\TybodyCase{$\odot_1$ }{$-$}{$M(.)$}
if $\odot_1=-$,\\
\TyFind{(- e_1)@x}
\TybodyProbe
\TyLHS{B}{3}{\genTTyA{\sigma}}
\spcC
\inferrule
    {
    \inferrule
        {\linegamF{e_1} \invNeg}
        {\linegamF{-e_1}\invProbe}
    \spcD \TyJud{\Gamma,{\sigma}}{x}{\TensorTy{d}}
    }
    {\linegamT{(- e_1)@x}}
\\
\TyLastFind{}
\TyFind{ -(e_1@x)}
\TyGiven{\linegamF{e_1}}{
 }
\\ \spcB
 then 
\TyApply{\judgeProbe}{ \linegamT{e_1@x}}
\\\spcB
and
\TyApply{\judgeNeg}{\linegamT{-e_1@x} }\\
T(\ruleNum{B}{3} \text{ for } $\odot_1=-$)  \\
if $\odot_1 = M(e_1)$\\
 \spcB Note: $M(e_1)=\sqrt{e_1}\mid \kappa (e_1) \mid \text{exp}(e_1) \mid  e^n$\\
\TyFind{(M (e_1))@x}
\TybodyProbe
\TyLHS{B}{3}{\genTTyA{\sigma}}
\spcC
\inferrule
    {
    \inferrule
        {\linegamF{e_1} (\invM)}
        {\linegamF{M(e_1)}\invProbe}
        \spcD \TyJud{\Gamma,{\sigma}}{x}{\TensorTy{d}}
    }
    {\linegamT{M(e_1)@x}}
\\
\TyLastFind{}
\TyGiven{\linegamF{e_1}}{}
\\ \spcB
then \TyApply{\judgeProbe}{ \linegamT{e_1@x}}
\\\spcB
and \TyApply{\judgeM}{\linegamT{M(e_1@x) }}\\
  \spcB T(\ruleNum{B}{3} \text{ for } $\odot_1=M$)  \\
\donecaseT{B}{3}
 \TyHead{B}{4}. Included in the earlier prose.
\TyHead{B}{5}\\
\TybodyCase{$\chi$}{$ \lift{e_1}$}{$\delta_{ij} \mid \mathcal{E}_\alpha$}
case  $\chi= \lift{e_1}$\\
\TyFind{  (\chi (e_1))@x}
\TybodyProbe
\TyLHS{B}{5}{\genTTyA{\sigma}}   
\spcC \inferrule
    {\inferrule
        {\linegamF{e_1} (\invLift)}
        {\TyEnv \vdash  (\lift{e_1}):\genFTyA{\sigma}\invProbe}
     \spcD \TyJud{\Gamma,{\sigma}}{x}{\TensorTy{d}}
    }
    {\TyEnv \vdash  ( \lift{e_1})@x:\genTTyA{\sigma}}
\\
\TyLastFind{}
\TyGiven{\linegamF{e_1}}{}
 \\ \spcB then
\TyApply{\judgeProbe}{ \linegamT{e_1@x}}
\\\spcB
and
\TyApply{\judgeLift}{\linegamT{\lift{e_1@x}}}\\
T(\ruleNum{B}{5}  \text{ where }$\chi= \lift{e_1}$)  \\
For the case $\chi=\delta_{ij} \mid \mathcal{E}_\alpha$\\
\TyGiven{\linegam{\chi}{\tau}}{ assumption }
then 
\TyApply{\judgeProbe}{ \linegam{\chi@x}{\tau}}\\
  \spcB T(\ruleNum{B}{5} \text{ where } $\chi= \delta_{ij} \mid \mathcal{E}_\alpha$)  \\
\donecaseT{B}{5}
\TyHead{C}{14}. Included in the earlier prose.
\TyHead{C}{11}.
Included in the earlier prose.
 \TyHead{C}{6}\\
 \TyFind{\ruleSrc{C}{6}}
\TybodyDerivS{\sqrt{e_1}}\\
\TyLHS{C}{6}{$\genFTyA{i}$} 
{\spcC
\inferruleLayer
    {\linegamS{e_1}     \invM}
    {\linegamS{\sqrt{e_1}} \invDeriv}
    { \linegamF{\ruleSrc{C}{6}} \text{ and } \sigma = \{i:d\} (\text{Claim})}
}\\
\TyLastFind{}
\TyGiven{ \linegamS{e_1}}{}\\
{\spcB then  $\linegamD{\einpart{i:d}  e_1} 
      ( \judgeDeriv) $}
\\\spcB
and 
{ $\linegamS{\sqrt{e_1}}
      (
       \judgeM) $}\\
\spcB Additionally, \TyApply{\judgeLift }{\linegamF{\lift{-} }}     \\ 
\TyGiven{\linegamS{\sqrt{e_1}} \text{ and }\linegamD{\einpart{i:d}  e_1}}{(\text{\ref{lne:derivr8} and \ref{lne:sqrtr8}})}
\\ \spcB then 
\TyApply{\judgeDiv}{\linegamD{\frac{\einpart{i:d}  e_1}{\sqrt{e_1}}}}      
 \\\spcB and 
\TyApply{\judgeProd}{\linegamD{\ruleTar{C}{6}}}
\TyHead{C}{7}\\
\TyFind{\ruleSrc{C}{7}}
\TybodyDerivS{ \textbf{cosine} (e_1)}\\
\TyLHS{C}{7}{$\genFTyA{i}$}
{\spcC 
\inferruleLayer
    { \linegamS{  e_1}\invM}
    {  \linegamS{  \textbf{cosine} (e_1)}}
    {\linegamD{\ruleSrc{C}{7}}  }
}\\
\TyLastFind{}
\TyGiven{\linegamS{  e_1}}{ }
\\ \spcB then
\TyApply{\judgeDeriv}{ \linegamD{\einpart{i}  e_1} },
\\\spcB
 \TyApply{\judgeM}{ \linegamS{\textbf{sine}(e_1)} },
\\ \spcB
\TyApply{\judgeNeg}{ \linegamS{ -\textbf{sine}(e_1)} }, 
\\\spcB and
\TyApply{\judgeProd}{\linegamD{(- \textbf{sine}(e_1))*(\einpart{i}  e_1)} }\\
\donecaseT{C}{7}
\TyHead{C}{8}. Included in the earlier prose.
\donecaseT{C}{8}
\TyHead{C}{18}\\
\TybodyDerivS{\textbf{tangent}  (e_1)}\\
\TyLHS{C}{18}{$\genFTyA{i}$}
{ \spcC 
\inferruleLayer
    { \linegamS{e_1} \invM }
    {  \linegamS{ \textbf{tangent} (e_1)}}
   {\linegamD{\ruleSrc{C}{18}}  }
   }\\
\TyLastFind{}
\TyGiven{\linegamS{ e_1}}{ }
\\ \spcB 
then \TyApply{\judgeDeriv}{ \linegamD{\einpart{i}  e_1} },
\\ \spcB
\TyApply{\judgeM, \judgeProd}{ \linegamS{ \textbf{cosine}(e_1) *\textbf{cosine}(e_1)}},
\\ \spcB and \TyApply{\judgeDiv}{ \linegamS{\ruleTar{C}{18} }}\\
\donecaseT{C}{18}
\sameRuleTyT{C}{9}{C}{10}
\TyHead{C}{10}\\
\TyFind{\ruleSrc{C}{10}}
\TybodyDerivS{ \textbf{arcsine}  (e_1)}\\
\TyLHS{C}{10}{$\genFTyA{i}$}
{ \spcC 
\inferruleLayer
    { \linegamS{  e_1}     ( \invM)}
    {  \linegamS{ \textbf{arcsine} (e_1)}}
    {\linegamD{\ruleSrc{C}{10}}  }
}\\
Since $\linegamS{  e_1}$   then 
\TyApply{\judgeDeriv }{\linegamD{\einpart{i}   e_1}}\\
\TyFind{\lift{1}}
{\spcC$ \TyEnv \vdash  \lift{1}:\genFTyA{\sigma}$($\judgeLift$)
}\\
\TyLastFind{}
\TyGiven{\linegamS{ e_1}}{.}
\\ \spcB
then \TyApply{\judgeProd}{\linegamS{ e_1*e_1}},
\\ \spcB
\TyApply{\judgeAdd}{\linegamS{ \lift{ 1}-(e_1*e_1)}},
\\\spcB
\TyApply{\judgeM}{\linegamS{\sqrt{ \lift{ 1}-(e_1*e_1)}}},
\\ \spcB
\TyApply{\judgeDiv}{\linegamS{\frac{ \lift{1}}{\sqrt{ \lift{1}-(e_1*e_1)}}}},
\\\spcB and
 \TyApply{\judgeProd}{\linegamD{\ruleTar{C}{10}}}\\
\donecaseT{C}{10}
\sameRuleTyT{C}{19}{C}{10}
\TyHead{C}{20}\\
\TyFind{\ruleSrc{C}{20}}
\TybodyDerivS{\exp  (e_1)}\\
\TyLHS{C}{20}{$\genFTyA{i}$}
{\spcC 
\inferrule
    {\linegamS{e_1} (\invM)}
    {\linegamS{\textbf{exp}(e_1)}
    }
}\\
\TyLastFind{}
\TyGiven{\linegamS{e_1}}{ }
\\\spcB
then \TyApply{\judgeDeriv}{ \linegamD{ \einpart{i} e_1}},
\\\spcB
\TyApply{\judgeM}{ \linegamS{\textbf{exp}(e_1)}},
\\\spcB  and 
\TyApply{\judgeProd}{ \linegamD{  \ruleTar{C}{20}}}\\
\donecaseT{C}{20}
\TyHead{C}{21}\\
\TybodyDerivS{ e_1^n}\\
\TyLHS{C}{21}{$\genFTyA{i}$}
\spcC{
    \inferruleLayer
    {\linegamS{e_1},\linegamR{n} \text{ and } \sigma =\{ i:d \}
     (\invM)
    }
    {\Gamma , \sigma  \setminus i  \vdash  (e^n):\genFTyA{\sigma  \setminus i }
       \invDeriv
    }
    { \TyEnv \vdash  \einpart{i:d} (e^n):\genFTyA{i}
    }
}\\
\TyLastFind{}
\TyGiven{\linegamS{e_1}}{}
then 
\TyApply{\judgeDeriv}{\linegamD{\einpart{i}  e_1}}.\\
\TyGiven{\linegamS{e_1}, \linegamR{n}}{ }
\\ 
\spcC then \TyApply{\judgeLift}{ \linegamS{\lift{n}}} and \TyApply{\judgeM}{ \linegamS{e^{n-1}}}.\\
\TyGiven{\linegamS{e^{n-1}} \text{ and } \linegamD{\einpart{i}  e_1}}{previous }
\\ \spcB
then \TyApply{\judgeProd}{ \linegamD{\ruleTar{C}{21}}}.\\
  \donecaseT{C}{21}
\TyHead{C}{16}\\
\TyFind {\einpart{i}(e_1 \odot e_2)}
\TybodyDeriv{e_1 \odot e_2}
\TyLHS{C}{16}{$\genFTyA{\sigma}$}
\TyFind{\tau(e_1) \text{ and }\tau(e_2)}
{\spcC
\inferruleLayer
    {\Gamma , \sigma  \setminus i  \vdash  e_1 , e_2:\genFTyA{\sigma  \setminus i }
     \invAdd
    }
    {\Gamma , \sigma  \setminus i  \vdash  e_1 \odot e_2:\genFTyA{\sigma  \setminus i }
       \invDeriv
    }
    {\linegamF{\einpart{i}(e_1 \odot e_2)}
    }
}\\
\TyLastFind{}
\TyGiven{\linegam{e_1,e_2}{\genFTyA{\sigma  \setminus i  }}}{Ln. \lineref{lne:e1r17} }
\\ \spcB
then \TyApply{\judgeDeriv}{\linegamF{\einpart{i}(e_1)}}
\\\spcB
and \TyApply{\invAdd}{\linegamF{\ruleTar{C}{16}}}.\\
\donecaseT{C}{16}
\TyHead{C}{15}\\
\TyFind {\einpart{i}(-e_1)}
\TybodyDeriv{-e_1}
\TyLHS{C}{15}{$\genFTyA{\sigma}$}
{\spcC
\inferruleLayer
    {\Gamma , \sigma  \setminus i  \vdash  e_1 :\genFTyA{\sigma  \setminus i }    
     \invNeg
    }
    {\Gamma , \sigma  \setminus i  \vdash  -e_1:\genFTyA{\sigma  \setminus i }
       \invDeriv
    }
    {\linegamF{\einpart{i}(-e_1)}
    }
}\\
\TyLastFind{}
\TyGiven{\linegam{e_1}{\genFTyA{\sigma  \setminus i  }}}{ }
\\ \spcB
then \TyApply{\judgeDeriv}{\linegamF{\einpart{i}(e_1)}}
\\\spcB
and \TyApply{\invNeg}{\linegamF{\ruleTar{C}{15}}}\\
\donecaseT{C}{15}
 \TyHead{C}{3}\\
 \TybodyDeriv{\sum\limits_{v=1}^{n}}
 \TyLHS{C}{3}{$\genFTyA{\sigma}$}
{\spcC
\inferruleLayer
    {\Gamma, {\sigma  \setminus i ,v:n} \vdash e_1:\genFTyA{\sigma  \setminus i,v:n }
     (\invSum)
    }
    {\Gamma , \sigma  \setminus i  \vdash  (\sum\limits_{v=1}^{n}  e_1):\genFTyA{\sigma  \setminus i }
       \invDeriv
    }
    {\linegamF{\einpart{i:d}(\sum\limits_{v=1}^{n}  e_1)}
    }
}\\
\TyLastFind{}
\TyGiven{\linegam{e_1}{\genFTyA{\sigma  \setminus i,v:n }}}{}
\\\spcB 
then \TyApply{\judgeDeriv}{\linegam{\einpart{i:d}(e_1)}{\genFTyA{\sigma,v:n }}}
\\\spcB
and \TyApply{(\judgeSum)}{\linegamF{  \sum\limits_{v=1}^{n} ( \einpart{i:d}(e_1))}}\\
\donecaseT{C}{3}
\TyHead{C}{2}\\
\TybodyDeriv{\nabla  \chi}
\spcB Lastly,\TyApply{\judgeLift}{\linegamF{\ruleTar{C}{2}}}.
\donecaseT{C}{2}
\TyHead{C}{5}\\
\spcB Given\TyApply{\judgeConv}{\linegam{V_\alpha \circledast H^{v}}{\genFTyA{\sigma /\ i }}}
\\\spcB
then \TyApply{\judgeDeriv}{\linegamF{\ruleSrc{C}{5}}}.
\\
\spcB
Lastly, \TyApply{\judgeConv}{\linegamF{\ruleTar{C}{5}}}.\\
\donecaseT{C}{5}
\TyHead{D}{8}\\
\TyFind{\ruleSrc{D}{8}}
\spcB \TyClaimGen{e_1}\\
{\spcC
    \inferruleLayer
        {\linegam{\ruleSrc{D}{8}}{\tau} \invNeg}
        {\linegam{-e_1}{\tau} \invNeg}
        {\linegam{\ruleTar{D}{8}}{\tau} }
}  \\
\TyLastFind{}
 \donecaseT{D}{8}
\TyHead{D}{1}\\
\TyFind {-0}
\spcB \TyClaimGen{-0}\\
\TyFind{0}
\TyLHS{D}{1}{$\tau$}
{\spcC
    \inferrule
    {\Gamma , \sigma \vdash 0:\tau \invNeg}
    {\Gamma , \sigma \vdash -0:\tau }
}\\
\TyLastFind{}
 \donecaseT{D}{1}
\TyHead{D}{3}\\
\TyFind {e_1 -0}
\spcB \TyClaimGen{e_1 -0}\\
\spcC $\linegamA{e-0} $\\
\spcC\TyApply{\judgeBase}{ \linegamA{0}}\\
\spcC T(\ruleNum{D}{3}) \\
  \donecaseT{D}{3}
\sameRuleTyT{D}{4}{D}{3}
\sameRuleTyT{D}{5}{D}{3}
\TyHead{E}{1}. Included in the earlier prose.
\sameRuleTyT{E}{2}{E}{1}
\TyHead{E}{4}\\
\TyLHS{E}{4}{$\einTyB{\sigma}$}
\spcC{\infer
    {
    \infer
        {\linegamA{e_1}\spcB
        \linegamAS{e_2}    
        \invDiv}
        {\linegamA{(\frac{e_1}{e_2})}}
   \spcD
        {\infer
       {\linegamAS{e_3,e_4} \invDiv}
        {\linegamAS{(\frac{e_3}{e_4})}\invDiv}}
    }{\linegamA{\frac{\frac{e_1}{e_2}}{\frac{e_3}{e_4}}}}}
\\
\TyLastFind{}
\TyFind{\frac{e_1 e_4}{e_2 e_3}}
\TyGiven{\linegamA{e_1} \text{ and } \linegamAS{e_2,e_3,e_4}}{ }
\\\spcB
then \TyApply{\judgeProd}{ \linegamA{ e_1*e_4}},
\\\spcB
\TyApply{\judgeProd}{ \linegamAS{ e_2*e_3}}{},
\\\spcB
 and \TyApply{\judgeDiv}{ \linegamA{ \frac{e_1 e_4}{ e_2 e_3}}}.\\
\donecaseT{E}{4}
\sameRuleTyT{D}{2}{D}{3}
\sameRuleTyT{D}{6}{D}{3}
\TyHead{E}{6}\\
\TyClaimGen{\ruleSrc{E}{6}}\\
\TyFind{e_1}
{ \spcC
    \inferruleLayer
    {\TyEnv \vdash e_1:\tau (\invM)}{\Gamma , \sigma     \vdash \sqrt{e_1}:\tau  \invProd}
    {\Gamma , \sigma \vdash \sqrt{e_1}*\sqrt{e_1}:\tau}
}\\
  \donecaseT{E}{6}
\sameRuleTyT{A}{4}{A}{3}
\TyHead{A}{3}\\
Given \TyApply{\judgeConv}{\linegamF{V_\alpha \circledast h^{jk}}}
\\
\spcB then \TyApply{\judgeProdEps}{\linegamF{\epsilon_{ijk} V_\alpha \circledast h^{jk}}}.\\
Lastly, \TyApply{\judgeLift}{\linegamF{\ruleTar{A}{3}}} \\
  \donecaseT{A}{3}
\TyHead{A}{1}\\
We know \TyApply{\judgeProdEps}{\linegamT{\ruleSrc{A}{1}}}.\\
Given
\TyApply{\judgeProdDel}{\linegamT{ \delta_{jl} \delta_{km} }
}
\\\spcB
 then \TyApply{\judgeAdd}{\linegamT{\ruleTar{A}{1} }\spc}.\\
\donecaseT{A}{1}
\TyHead{A}{5}\\  
\TyFind {\ruleSrc{A}{5}}
Given \TyApply{  \judgeBase}{\linegam{T_j }{ \genTTyA{\sigma}} \text{ and }\sigma =\{j\}}\\
 \spcB then \TyApply{ \judgeProdDel}{\linegam{ \delta_{ij} (T_j)}{ \genTTyA{\sigma}}}
 \\ \spcB
 $\text{and }\sigma =\{i \}\judgeProdDel $ \\
 \TyFind {\ruleTar{A}{5}}
\spcC{  
     {$\linegam{T_i}{ \genFTyA{\sigma }} \text{ and }\sigma =\{i \}\judgeBase$
     }    
 }\\
   \donecaseT{A}{5}
\sameRuleTyT{A}{6}{A}{5}
\TyHead{A}{8}\\
Given \TyApply{ \judgeConv}{ \linegam{V \circledast H^{\delta_{cj}} }{ \genFTyA{\sigma}} \text{ and }\sigma =\{j\}}    \\
\spcB then \TyApply{\judgeProdDel}{\linegam{ \delta_{ij} (V \circledast H^{\delta_{cj}})}{ \genFTyA{\sigma}}\text{ and }\sigma =\{i \}}\\
\spcC  $\linegam{V \circledast H^{\delta_{ci}} }{ \genFTyA{\sigma }} \text{ and }\sigma =\{i \} \judgeConv$ \\
   \donecaseT{A}{8}
\sameRuleTyT{A}{9}{A}{8}
\TyHead{A}{7}. Included in the earlier prose.
\TyHead{E}{5}. Included in the earlier prose.
\TyHead{C}{22}\\
This type of structure inside a derivative operation results in a field type. \\
Claim: $\Gamma, {\sigma /\  \alpha \beta} \vdash  e_1:\genFTyA{\sigma /\  \alpha \beta}$ 
\\ 
\TyLHS{C}{22}{$\genFTyA{\sigma}$}
\inferruleLayer
{\linegam{    e_1}{\genFTyA{\sigma /\ \alpha \beta} }\judgeDeriv}
{\linegam{ (\einpart{\beta}   e_1)}{\genFTyA{\sigma /\ \alpha} }\judgeDeriv }
{\linegamF{( \einpart{\alpha}  \einpart{\beta}   e_1)}}\\
\TyLastFind{}
\TyGiven{\linegam{ e}{\sigma /\ \alpha \beta}}{-}
then \TyApply{\judgeDeriv}{\linegamF{ \einpart{\beta \alpha}e}}\\
 \donecaseT{C}{22}
 T(d) Lemma \ref{thm:type}
\section{Value Preservation Proof}
 \label{proof:valuePreserv}

The following is a proof for Theorem \ref{thm:value}
 Given a derivation $d$ of the form $e \longrightarrow e'$ we state \thmNameV(d) as a shorthand for the claim that the derivation preserves 
the value of the expression $e$.
The proof demonstrates that $\forall d. V(d)$.
\newline
\noindent Case on structure of d\\
\textbf{Case }Rules R1-R5 use the probe operator.\\
\spcB Value representation of the probe operator is not supported.\\
\textbf{Case }Rules R6-R21 use the differentiation operator.\\
\spcB Value representation of the differentiation operator is not supported.
\TyHead{D}{8}\\
$\valHeadingR{D}{8}$
\\ \spc\spc 
then 
\VApply{\valunary}{-e_1  \Downarrow -v'  }, 
\\ \spc\spc
and 
\VApply{\valunary}{--e_1  \Downarrow --v'}\\
The value of $v $ is   $--v'$.\\ 
\spc\spc By using algebraic reasoning:  $--v' = v'$.
 Since $--e_1  \Downarrow v \text{ and } --e_1  \Downarrow v' \text{ then }  v=v'$.\\
$\valEndProse{D}{8}$
\TyHead{D}{1}\\
$\valHeading{D}{1}$
\spc\spc then 
\VApply{\valbaseC}{0 \Downarrow \valReal{}(0)}
, and 
\VApply{\valunary}{  -0  \Downarrow  \valReal{}(-0)}\\
\spc
The value of $v $ is    $\valReal{}(-0)$\\ 
    \spc\spc  By using algebraic reasoning: $\valReal{}(-0) = \valReal{}(0)$\\
$\valEndProse{D}{1}$
\TyHead{D}{3}\\
\spcC Included in the earlier prose.
\TyHead{D}{4}\\
$\valHeadingR{D}{4}$
\\ \spc\spc 
then 
\VApply{(\valbaseC,\valbinary)}{0-e_1  \Downarrow \valReal{}(0)+v'}.\\
\spc
The value of $v $ is  $\valReal{}(0)+v' $.
By using algebraic reasoning: $\valReal{}(0)+v'  = v'$.
\\ \spc\spc 
Since $0-e_1  \Downarrow v \text{ and }0-e_1  \Downarrow v' \text{ then }  v=v'$\\
$\valEndProse{D}{4}$
\TyHead{D}{5}\\
\spc Assume that $e_1  \Downarrow \valReal{}(v2)$
then 
\VApply{(\valbaseC,\valbinary)}{\frac{0}{e_1}  \Downarrow \valReal{}(\frac{0}{v2})}.\\
\spc
The value of $v $ is   $\valReal{}(\frac{0}{v2}) $.
 By using algebraic reasoning: $\valReal{}(\frac{0}{v2})=\valReal{}(0)$ \\
\spc Lastly, 
\VApply{(\valbaseC{})}{0 \Downarrow \valReal{}(0) }\\
$\valEndProse{D}{5}$
\TyHead{E}{1}\\ 
$\valHeadingR{E}{1}, e_1 \Downarrow v1,e_2 \Downarrow v2,e_3 \Downarrow v3$.
\\ \spc\spc
then 
\VApply{\valbinary}{\frac{e_1 }{ e_2} \Downarrow \frac{v1}{ v2} }
and 
\VApply{\valbinary}{\frac{\frac{e1}{e_2 }}{e_3} \Downarrow   \frac{\frac{v1}{v2}}{ v3} }.
\\ 
\TyGiven{e_1 \Downarrow v1 \spc e_2  \Downarrow v2  \spc e_3  \Downarrow v3}{}
\\ \spc\spc then 
\VApply{\valbinary}{{e_2 }{ e_3} \Downarrow {v2}*{ v3}}
and 
\VApply{\valbinary}{\frac{e1}{e_2e_3} \Downarrow   \frac{v1}{v2* v3}}.\\
\spc
The value of $v $ is  $\frac{v1}{v2* v3}$.
By using algebraic reasoning:
 $ v'=\frac{v1}{v2 * v3} =\frac{\frac{v1}{v2}}{ v3}=v$.\\
$\valEndProse{E}{1}$
\sameRuleTyV{E}{2}{E}{1}
\sameRuleTyV{E}{4}{E}{1}
\TyHead{D}{2}
$\valHeadingR{D}{2}$
then 
\VApply{(\valbaseC, \valbinary{})}{e_1+0  \Downarrow v' +\valReal{}(0)}. \\ 
\spc By using algebraic reasoning $v' +\valReal{}(0) = v'$\\
$\valEndProse{D}{2}$
\sameRuleTyV{D}{6}{D}{5}
\TyHead{E}{6}\\
\spcC Included in the earlier prose.
\valNo{A}{4}
\valNo{A}{3}
\TyHead{A}{1}\\
\spcC Included in the earlier prose.
\TyHead{A}{5}\\ 
\spcC Included in the earlier prose.\\
\textbf{Case }Rules R37-R40 uses field terms\\
\spcB Value representation of the field terms is not supported.\\
\TyHead{E}{5}\\
$\valHeading{E}{5}$ 
\spc
Assume that $ s \Downarrow v_s  $ and $ e_1 \Downarrow v_e $ 
\\ \spc\spc 
then \VApply{(\valbinary )}{ s* e_1 \Downarrow v_s* v_e  }
\\ \spc\spc 
and
\VApply{\valunary}{\ruleSrc{E}{5} \Downarrow \sum (v_s* v_e)}\\
\spc
The value of $v $ is $   \sum (v_s* v_e) $\\  
\spc\spc 
\TyBy{\text{moving scalar outside summation}}{v}{v_s * \sum ( v_e)}\\
 $\valTarM{E}{5}$
\TyGiven{  s \Downarrow v_s  \text{ and }e \Downarrow v_e }{}
\\ \spc \spc then
\VApply{ ( \valunary)}{ \sum  e \Downarrow \sum v_e}
and
\VApply{ ( \valbinary)}{ \ruleTar{E}{5} \Downarrow v_s *\sum v_e}\\
$\valEndProse{E}{5}$
     \valNo{C}{22}

\section{Termination}
\label{proof:termination}
\subsection{Size reduction}
\label{proof:termination1}
\textbf{
  If e $\Longrightarrow$ e' then $\mathcal{S}(e) > \mathcal{S}(e') \geq 0$ }
 (Lemma \ref{lem:size}).
 The following are a few helpful lemmas 
 that will be referred to in the proof.

\begin{lemma}  $5^{(1+x)}>(16+5^x) $\\
\label{lemx}
$\begin{array}{ll}
5^x>4  . &\text{Given }  x>=1\\
4*5^x> 16  &\text{Multiply by 4}\\
5*5^x- 5^x> 16 & \text{Refactor left side}\\
5*5^x>(16+5^x) & \text{Add }5^x \\
5^{(1+x)}>(16+5^x) & \text{Rewritten}\\
\end{array}$
\end{lemma}
\begin{lemma} $5^{(\sizea+\sizeb)}> 5^{(\sizea)}> 4$.\label{lemy}
\end{lemma}
\begin{lemma}
$ (1+\sizea)5^{(1+\sizea)}>\sizea(16+5^{\sizea} ) +20 $\\
\label{lemz}

$\begin{array}{lll}

5^{(1+\sizea)}>16+5^{\sizea} & Lemma ~\ref{lemx} \\
\sizea 5^{(1+\sizea)}>\sizea (16+5^{\sizea}) & \text{Multiply by } \sizea\\
\sizea 5^{(1+\sizea)}+5^{(1+\sizea)} 
 >\sizea (16 +5^{\sizea}) +5^{(1+\sizea)} & \text{Add }5^{(1+\sizea)}\\
(1+\sizea) 5^{(1+\sizea)}
 > \sizea (16 +5^{\sizea}) +5*5^{\sizea}>\sizea (16 +5^{\sizea}) +20 & (Lemma ~\ref{lemy}) 

\end{array}$
\end{lemma}

The following is a proof for Lemma ~\ref{lem:size}
Given a derivation $d$ of the form $e \longrightarrow e'$ we state \thmNameSize(d) as a shorthand for the claim that the derivation reduces the size of the expression $e$.
By case analysis and comparing the size metric provided.
This proof does a case analysis to show
$\forall d \in Deriv.  \thmNameSize(d)$.
 Case on structure of d\\
 \TyHead{B}{1}.
Included in the earlier prose.
\thmTALong{B}{2}{2+2\sizea+2\sizeb}{1+2\sizea +2\sizeb}{}
\thmTAShort{B}{3}{2+2\sizea}{1+2\sizea}{}\\
\spcB P(d) 
\thmTALong{B}{4}{4+4\sizea}{2+4\sizea}{}
\thmTAShort{B}{5}{2 \size(\chi)}{\size(\chi)}{}   
\TyHead{C}{14}\\
\spcB We define $\sizeof{(\ruleSrc{C}{14})}$={$s_1+s_2+s_3$}\\
\spcC where $s_1 =\sizea*5^{1+\sizea +\sizeb}$,
$ s_2 =\sizeb*5^{1+\sizea +\sizeb},$ and $ s_3=5^{1+\sizea +\sizeb},$\\
\spcB We define $\sizeof{(\ruleTar{C}{14})}$={$t_1+t_2+t_3$}\\
\spcC where $ t_1 =\sizea(5^{\sizea}+1) $,
$ t_2=\sizeb(5^{\sizea}+1),$ and $t_3=3$\\
\spcB Given $4*5^{1+\sizea} >1$ then \\
\spcC 
\TyApply{\text{adding } 5^{\sizea}}{\longrightarrow 5*5^{\sizea} >  5^{\sizea}+1} \\
\spcC 
\TyApply{\text{refactoring}}{\longrightarrow5^{1+\sizea +\sizeb} >  5^{\sizea}+1}  \\
\spcC 
\TyApply{\text{multiplying by }\sizea}{\longrightarrow \sizea*5^{1+\sizea +\sizeb}>  \sizea(5^{\sizea}+1)}
\\ \spcC
\TyApply{\text{ multiplying by }\sizeb}{\longrightarrow \sizeb*5^{1+\sizea +\sizeb}>  \sizeb(5^{\sizea}+1)}
\\
\spcC where and so $s_1> t_1$,$s_2> t_2$\\
\spcC where Lastly, $5^{1+\sizea +\sizeb}>3  $  (Lm ~\ref{lemy})  and so $s_3> t_3$\\  
\spcB Finally, $ \sizeof{\ruleSrc{C}{14}}> \sizeof{\ruleTar{C}{14}}$ \\
\spcB P(d) 
  \TyHead{C}{11}\\
\spcB We define $\sizeof{(\ruleSrc{C}{11})}$={$s_1+s_2+s_3$}\\
\spcC where $s_1 =\sizea 5^{2+\sizea+\sizeb}$, $s_2 =\sizeb 5^{2+\sizea+\sizeb}$
, and $s_3=2*5^{2+\sizea+\sizeb}$\\
\spcB We define $\sizeof{(\ruleTar{C}{11})}$={$t_1+t_2+t_3$}\\
\spcC where $t_1 =\sizea(1+5^{\sizea})$
, $t_2 = \sizeb(3+5^{\sizeb})$, and $t_3 =6$\\
\spcB Given $ 5^{2+\sizea+\sizeb}>(1+5^{\sizea}) $(Lm ~\ref{lemx})\\
\spcC where 
then \TyApply{\text{multiplying by }\sizea}{\sizea 5^{2+\sizea+\sizeb}>\sizea(1+5^{\sizea})}
\\\
 \spcC where so $s_1> t_1$,$s_2> t_2$\\
  \spcB Given $5^{1+\sizea+\sizeb}> 5^{\sizeb} +3$  (Lm ~\ref{lemx})\\
 \spcC where then \TyApply{\text{multiplying by }2}{ 2*5^{1+\sizea+\sizeb}> 2*5^{\sizeb}+6}\\
 \spcC where so $s_3> t_3$\\ 
\thmTlastline{}
\thmTALong{C}{6}
    {(1+\sizea)5^{(1+\sizea)}}
    {\sizea(1+ 5^{\sizea})+6}
    {(Lm ~\ref{lemz})}
\TyHead{C}{7}. Included in the earlier prose.
\thmTALong{C}{8} 
    {(1+\sizea)5^{(1+\sizea)}}
    {\sizea(1+5^{\sizea} ) +2 }
    {(Lm ~\ref{lemz})}
\thmTALong{C}{18}
    {(1+\sizea)5^{(1+\sizea)}}
    {{\sizea}(5^{\sizea}+2)+5}
    {(Lm ~\ref{lemz})}
\thmTALong{C}{9}
    {(1+\sizea)5^{(1+\sizea)}}
    {\sizea(2+5^{\sizea}) +11}
    { (Lm ~\ref{lemz})}
\thmTALong{C}{10}
    {(1+\sizea)5^{(1+\sizea)}}
    {\sizea(2+5^{\sizea}) +10}
    {(Lm ~\ref{lemz})}
\thmTALong{C}{19}{(1+\sizea)5^{(1+\sizea)}}
{{\sizea}(2+5^{\sizea})+9}
{(Lm ~\ref{lemz})}
\thmTALong{C}{20}
    {(1+\sizea)5^{(1+\sizea)}}
    {\sizea(1+5^{\sizea})+2}
    {(Lm ~\ref{lemz})}
\thmTALong{C}{21}
    {(1+\sizea)5^{(1+\sizea)}}
    {5+\sizea(1+5^{\sizea})}
{(Lm ~\ref{lemz})}
\TyHead{C}{16}Included in the earlier prose.
\thmTALong{C}{15}{5^{1+\sizea}(1+\sizea)}{1+\sizea5^\sizea}{(Lm ~\ref{lemx})}
\thmTALong{C}{3}{(2+2\sizea)*5^{2+2\sizea}}{2+2\sizea 5^{\sizea}}{}
\thmTAShort{C}{2}{{\size{\chi}}5^{\size{\chi}}}{2}{(Lm ~\ref{lemy})}
\thmTAShort{C}{5}{5}{1}{}
\thmTAShort{D}{8}{2+\sizea}{\sizea}{}
\thmTAShort{D}{1}{2}{1}{}
\thmTAShort{D}{3}{2+\sizea}{\sizea}{}
\sameRuleTyP{D}{4}{D}{3}
\thmTAShort{D}{5}{3+\sizea}{1}{}
\TyHead{E}{1}\spcC Included in the earlier prose.
\sameRuleTyP{E}{2}{E}{1}
\thmTAShort{E}{4}
    {6+\sizeof{e_1}+\sizeb+\sizeof{e_3}}
    {4+\sizeof{e_1}+\sizeb+\sizeof{e_3}}{}
\thmTAShort{D}{2}{2+\sizea}{\sizea}{}
\sameRuleTyP{D}{6}{D}{2}
\thmTAShort{E}{6}{3+2\sizea}{\sizea}{}
\thmTAShort{A}{4}{5+\sizea 5^\sizea}{2}{}
\thmTAShort{A}{3}{6}{2}{}
\thmTAShort{A}{1}{9}{7}{}
\thmTAShort{A}{5}{3}{1}{}
\sameRuleTyP{A}{6}{A}{5}
\sameRuleTyP{A}{8}{A}{5}
\thmTAShort{A}{9}{4}{2}{}
\thmTAShort{A}{7}{2+\sizea 5^{\sizea}}{\sizea 5^\sizea}{}
\thmTAShort{E}{5}{6+2\sizea}{4+2\sizea}{} \\
\spcA \thmNameSize(d) Lemma \ref{lem:size}

\subsection{Termination implies Normal Form}
\label{proof:termination2}
\textbf{Termination implies normal form} (Lemma \ref{lem2}).
  The proof is by examination of the EIN syntax in [2].
  For any syntactic construct, we show that either the term is in normal form, or there
  is a rewrite rule that applies (\secref{proof:termination2}).
We state $Q(e_x)$ as a shorthand for the claim that if $x$  has terminated and is normal form.
Additionally we state $CQ(e_x)$ if there exists an expression that is not in normal form and has terminated.
The following is a proof by contradiction. \\
Define the following shorthand:
M($e_1$) = $\sqrt{e_1} \mid exp(e_1)\mid e_1^n \mid \kappa(e_1) $\\
\noindent case  on structure $e_x$ \\
$\begin{array}{llll}
\normalLine{x}{c}
\normalLine{x}{T_\alpha}
\normalLine{x}{F_\alpha}
\normalLine{x}{V_\alpha \circledast H}
\normalLine{x}{\delta_{ij}}
\normalLine{x}{\mathcal{E}_\alpha}
&\text{If }e_x=\lift{e_1}  \\
\end{array}$\\
\spcB Prove \thmNameN($e_x$) by contradiction.\\
\spcB case on structure $e_1$\\ 
$\begin{array}{lll}
\spcC
\normalLine{1}{c}
\normalLine{1}{T_\alpha}
\normalTy{1}{F_\alpha}
\normalTy{1}{e \circledast e}
\normalLine{1}{\delta_{ij}}
\normalLine{1}{\mathcal{E}_\alpha}
\normalTy{1}{\lift{e}}
\normalNA{1}{M(e)}
& &\text{Given }M(e_3) = \sqrt{e_3} \mid exp(e_3)\mid e_3^n \mid \kappa(e_3) \\
\normalNA{1}{-e}
\normalTy{1}{\einpart{\alpha}e}
\normalNA{1}{\sum e}
\normalNAS{1}{e_3+e_4}
\normalNAS{1}{e_3-e_4}
\normalNAS{1}{e_3*e_4}
\normalNAS{1}{\frac{e_3}{e_4}}
\normalNAS{1}{e_3@e_4}
& Q(e_x)
\end{array}$ \\
\thmNPros{M($e_1$)}{\thmN{G}{5}}{$e_1$} 
\spcB Note. M($e_1$) = $\sqrt{e_3} \mid exp(e_3)\mid e_3^n \mid \kappa(e_3) $\\
$\begin{array}{lll}
\spcC
\normalLine{1}{c}
\normalLine{1}{T_\alpha}
\normalLine{1}{F_\alpha}
\normalLine{1}{V_\alpha \circledast H}
\normalLine{1}{\delta_{ij}}
\normalLine{1}{\mathcal{E}_\alpha}
\normalNA{1}{\lift{e}}
\normalNA{1}{M(e)}
\normalNA{1}{-{e}}
\normalNA{1}{\frac{\partial}{\partial x_\alpha}{e}}
\normalNA{1}{\sum{e}}
\normalNAS{1}{e_3+e_4}
\normalNAS{1}{e_3-e_4}
\normalNAS{1}{e_3*e_4}
\normalNAS{1}{\frac{e_3}{e_4}}
\normalNAS{1}{e_3@e_4}
& Q(e_x)
\end{array}$ \\
\thmNPros{$-e_1$}{\thmN{G}{5}}{$e_1$} 
$\begin{array}{lll}
\spcC
\normalRule{1}{0}{\ruleNum{D}{1}}
\normalLine{1}{c}
\normalLine{1}{T_\alpha}
\normalLine{1}{F_\alpha}
\normalLine{1}{V_\alpha \circledast H}
\normalLine{1}{\delta_{ij}}
\normalLine{1}{\mathcal{E}_\alpha}
\normalNA{1}{\lift{e}}
\normalNA{1}{M(e)}
\normalRule{1}{-{e}}{\ruleNum{D}{8}}
\normalNA{1}{\frac{\partial}{\partial x_\alpha}{e}}
\normalNA{1}{\sum{e}}
\normalNAS{1}{e_3+e_4}
\normalNAS{1}{e_3-e_4}
\normalNAS{1}{e_3*e_4}
\normalNAS{1}{\frac{e_3}{e_4}}
\normalNAS{1}{e_3@e_4}
& Q(e_x)
\end{array}$ \\
$e_x=e_1+e_2$\\
\spcB Prove Q(x)\\
\spcB case on structure $e_1$\\ 
$\begin{array}{lll}
\spcC
\normalRule{x}{0}{R30}
\normalLine{x}{c}
\normalLine{x}{T_\alpha}
\normalLine{x}{F_\alpha}
\normalLine{x}{V_\alpha \circledast H}
\normalLine{x}{\delta_{ij}}
\normalLine{x}{\mathcal{E}_\alpha}
\normalNA{x}{\lift{e}}
\normalNA{x}{M(e)}
\normalNA{x}{-{e}}
\normalNA{x}{\frac{\partial}{\partial x_\alpha}{e}}
\normalNA{1}{\sum{e}}
\normalNAS{1}{e_3+e_4}
\normalNAS{1}{e_3-e_4}
\normalNAS{1}{e_3*e_4}
\normalNAS{1}{\frac{e_3}{e_4}}
\normalNAS{1}{e_3@e_4}
& Q(e_x)
\end{array}$ \\
\spcB case on structure $e_2$\\
\spcC Proof same as above\
\spcC Q(x)\\
\thmNPros{$e_1-e_2$}{\thmN{G}{5}}{$e_1$} 
$\begin{array}{lll}
\spcC
\normalRule{1}{0}{R25}
\normalLine{1}{c}
\normalLine{1}{T_\alpha}
\normalLine{1}{F_\alpha}
\normalLine{1}{V_\alpha \circledast H}
\normalLine{1}{\delta_{ij}}
\normalLine{1}{\mathcal{E}_\alpha}
\normalNA{1}{\lift{e}}
\normalNA{1}{M(e)}
\normalNA{1}{-{e}}
\normalNA{1}{\frac{\partial}{\partial x_\alpha}{e}}
\normalNA{1}{\sum{e}}
\normalNAS{1}{e_3+e_4}
\normalNAS{1}{e_3-e_4}
\normalNAS{1}{e_3*e_4}
\normalNAS{1}{\frac{e_3}{e_4}}
\normalNAS{1}{e_3@e_4}
& Q(e_x)
\end{array}$ \\
\spcB case on structure $e_2$\\
$\begin{array}{lll}
\spcC
\normalRule{x}{0}{R24}
& \text{Proof same as above}\\
& Q(x) \\
\end{array}$ \\
\thmNPros{$e_1*e_2$}{\thmN{G}{5}}{$e_1$} 
$\begin{array}{lll}
\spcC
\normalRule{1}{0}{R31}
\normalLine{1}{c}
\normalLine{1}{T_\alpha}
\normalLine{1}{F_\alpha}
\normalLine{1}{V_\alpha \circledast H}
\end{array}$ \\
\spcC \text{ If }$e_1= \delta_{ij}$\\
\spcE case on structure $e_2$\\
$\begin{array}{lll}
\spcE 
\normalRule{2}{T_{j}}{R36}
\normalRule{2}{F_{j}}{R37}
\normalRule{2}{V_\alpha \circledast H}{R38}
\normalRule{2}{V_\alpha \circledast H@e}{R39}
\normalRule{2}{\frac{\partial}{\partial x_\alpha}{e}}{R40}
\end{array}$ \\
\spcE else $Q(e_x)$ because $e_x$ is in normal form.\\
\spcC \text{If }$e_1= \mathcal{E}_{ij}$\\
\spcC If $e_1= \mathcal{E}_{ijk}$\\
$\begin{array}{lll}
\spcE 
&\text{case on structure }e_2\\
\normalRule{2}{\frac{\partial}{\partial x_{ij}}(e)}{R33}
\normalRule{2}{V \circledast H_{jk}}{R34}
\normalRule{2}{\mathcal{E}_{ijk}}{R35}
\end{array}$ \\
\spcE else $Q(e_x)$ because $e_x$ is in normal form.\\
$\begin{array}{lll}
\spcC \normalNA{1}{\lift{e_1}}
\spcC \text{If } e_1= \sqrt{e_3}\\
\end{array}$ \\
$\begin{array}{lll}
\spcE \normalRule{2}{\sqrt{e_4}}{R32}
\end{array}$ \\
\spcE otherwise $Q(e_x)$ because $e_x$ is in normal form.\\
$\begin{array}{lll}
\normalNA{1}{-e}
\normalTy{1}{\einpart{\alpha}e}
\normalNA{1}{\sum e}
\normalNAS{1}{e_3+e_4}
\normalNAS{1}{e_3-e_4}
\normalNAS{1}{e_3*e_4}
\normalNAS{1}{\frac{e_3}{e_4}}
\normalNAS{1}{e_3@e_4}
& Q(e_x)
\end{array}$ \\
\thmNPros{$\frac{e_1}{e_2}$}{\thmN{G}{5}}{$e_1$} 
\spcC If $e_1= \frac{e_3}{e_4}$\\
$\begin{array}{lll}
 \spcE  \normalRule{2}{\frac{e_5}{e_6}}{R27}
 \end{array}$ \\
\spcE otherwise $Q(e_x)$ because we can apply rule R29.\\
$\begin{array}{lll}
\spcC
\normalRule{1}{0}{R26}
\normalLine{1}{c}
\normalLine{1}{T_\alpha}
\normalLine{1}{F_\alpha}
\normalNA{1}{V \circledast H}
\normalLine{1}{\delta_{ij},\mathcal{E}_{ij},\mathcal{E}_{ijk}}
\normalNA{1}{\frac{\partial}{\partial x_\alpha}  e}
\normalNA{1}{\sum e }
\normalNA{1}{\lift{ e} }
\normalNA{1}{M(e)}
\normalNA{1}{-e}
\normalNA{1}{e+e}
\normalNA{1}{e-e}
\normalNA{1}{e*e}
\normalNA{1}{e@e}
\end{array}$\\
\spcB case on structure $e_2$\\
$\begin{array}{lll}
\spcC \normalRule{2}{\frac{e_4}{e_5}}{R28}
\end{array}$ \\
\spcC otherwise proof same as above\\
\spcB $Q(e_x)$ \\
\thmNPros{$e_1@e_2$}{\thmN{G}{5}}{$e_1$} 
$\begin{array}{lll}
\spcC
\normalTy{1}{c}
\normalTy{1}{T_\alpha}
\normalNA{1}{F_\alpha}
\normalNA{1}{e \circledast e}
\normalRule{1}{\delta_{ij},\mathcal{E}_\alpha}{R5}
\normalRule{1}{\lift{e}}{R5}
\normalRule{1}{M(e)}{R3}
\normalRule{1}{-e}{R3}
\normalNA{x}{\einpart{\alpha}e}
\normalRule{1}{\sum e}{R4}
\normalRule{1}{e+e}{R2}
\normalRule{1}{e-e}{R2}
\normalRule{1}{e*e}{R1}
\normalRule{1}{\frac{e}{e}}{R1}
\normalTy{1}{e@e}
& Q(e_x)
\end{array}$ \\
\thmNPros{$\frac{\partial}{\partial x_\alpha}  e_1$}{\thmN{G}{5}}{$e_1$} 
$\begin{array}{lll}
\spcC
\normalTy{1}{c}
\normalTy{1}{T_\alpha}
\normalLine{1}{F_\alpha}
\normalRule{1}{e \circledast e}{R21}
\normalRule{1}{\delta_{ij},\mathcal{E}_\alpha}{R20}
\normalRule{1}{\lift{e}}{R20}
\end{array}$ \\
\spcC \text{ If }$e_1 = M(e_2)$\\
\spcC \text{ case on structure }$e_2$\\
$\begin{array}{lll}
\spcE
\normalRule{2}{Cosine(e)}{R9}
\normalRule{2}{Sine(e)}{R10}
\normalRule{2}{Tangent(e)}{R11}
\normalRule{2}{ArcCosine(e)}{R12}
\normalRule{2}{ArcSine(e)}{R13}
\normalRule{2}{ArcTangent(e)}{R14}
\normalRule{2}{exp(e)}{R15}
\normalRule{2}{e^n}{R16}
\normalRule{2}{\sqrt{e}}{R8}
& Q(e_x)
\end{array}$ \\
$\begin{array}{lll}
\spcC
\normalRule{1}{-e}{R18}
\normalRule{1}{\einpart{\alpha}e}{R42}
\normalRule{1}{\sum e}{R19}
\normalRule{1}{e+e}{R17}
\normalRule{1}{e-e}{R17}
\normalRule{1}{e*e}{R6}
\normalRule{1}{\frac{e}{e}}{R7}
\normalTy{1}{e@e}
& Q(e_x)
\end{array}$ \\
\thmNPros{$\sum(e_1)$}{\thmN{G}{5}}{$e_1$} 
$\begin{array}{lll}
\spcC
\normalRule{1}{c}{R41}
\normalRule{1}{T}{R41}
\normalLine{1}{T_\alpha}
\normalRule{1}{F}{R41}
\normalLine{1}{F_\alpha}
\normalRule{1}{V_\alpha \circledast H}{R41}
\normalLine{1}{\delta_{ij},\mathcal{E}_\alpha}
\normalNA{1}{\lift{e}}
\normalNA{1}{M(e)}
\normalNA{1}{-{e}}
\normalNA{1}{\frac{\partial}{\partial x_\alpha}{e}}
\normalNA{1}{\sum{e_1}}
\normalNAS{1}{e_3+e_4}
\normalNAS{1}{e_3-e_4}
\normalNAS{1}{e_3*e_4}
\normalNAS{1}{\frac{e_3}{e_4}}
\normalRule{1}{F@e}{R41}
\normalRule{1}{V\circledast h@e}{R41}
\normalLine{1}{e@e}
& Q(e_x)
\end{array}$ \\

\subsection{Normal Form implies Termination}
\label{proof:termination3}
 The section offers a  proof for  Lemma \ref{lem3}. \\
\begin{definition} [Non-terminated]
A term has not terminated if it is 
the source term of a rewrite rule.  \end{definition}
\textbf{Normal form implies Termination.} (Lemma \ref{lem3}).
\begin{proof}
We state \thmNameL(e) as a shorthand for the claim that if $e$ is in normal form then it has terminated.
The following is a proof by contradiction. 
C\thmNameL(e): There exists an expression e that has not terminated  
and is in normal form.
More precisely, given a derivation $d$ of the form $e \longrightarrow e'$ ,
there exists an expression that is the source term $e$ of derivation  $d$ therefore not-terminated, and is in normal form. 
\end{proof} 
Case analysis on the source of each rule\\\thmL{B}{1}\thmL{B}{2}\thmL{B}{3}\thmL{B}{4}\thmL{B}{5}\thmL{C}{14}\thmL{C}{11}\thmL{C}{6}\thmL{C}{7}\thmL{C}{8}\thmL{C}{18}\thmL{C}{9}\thmL{C}{10}\thmL{C}{19}\thmL{C}{20}\thmL{C}{21}\thmL{C}{16}\thmL{C}{15}\thmL{C}{3}
\thmL{C}{1}\thmL{C}{2}\thmL{C}{5}\thmL{D}{8}\thmL{D}{1}\thmL{D}{3}\thmL{D}{4}\thmL{D}{5}\thmL{E}{1}\thmL{E}{2}\thmL{E}{4}\thmL{D}{2}\thmL{D}{6}\thmL{E}{6}\thmL{A}{4}\thmL{A}{3}\thmL{A}{1}\thmL{A}{5}\thmL{A}{6}\thmL{A}{8}\thmL{A}{9}\thmL{A}{7}\thmL{E}{5}\thmL{C}{22}\\
\spcA M(x) Lemma \ref{lem3} \\

\end{document}